\documentclass{elsarticle}
\usepackage[utf8]{inputenc}
\usepackage{ulem}
\usepackage{makeidx}
\usepackage{subcaption}
\usepackage[english]{babel}
\usepackage{graphicx}
\usepackage{amsmath,amsfonts,amssymb}
\usepackage{amstext}
\usepackage[mathscr]{eucal}
\usepackage{bm}
\usepackage{url}
\usepackage{pifont}
\usepackage{calc}
\usepackage{float}
\usepackage{latexsym}
\usepackage{paralist}
\usepackage{xspace}
\usepackage{cancel}
\usepackage{xcolor}
\usepackage{lineno}
\usepackage{geometry}
\usepackage{caption}
\usepackage{array}
\usepackage{multirow}
\DeclareGraphicsExtensions{.eps,.jpg,.png,.pdf}
\usepackage{amstext}
\usepackage{pifont}
\usepackage{colortbl}
\usepackage[pagebackref=false,bookmarks=false]{hyperref}
\usepackage{appendix}
\usepackage{float}
\usepackage{graphics}
\usepackage{subcaption}
\usepackage{xcolor}
\usepackage{xspace}
\usepackage{hyperref}
\hypersetup{
    colorlinks=false,
    linktoc=all
}
\usepackage{appendix}

\usepackage{enumitem}
\usepackage{longtable}
\usepackage{todonotes}
\usepackage{subcaption}

\usepackage{multirow,tabularx}

\addto\captionsenglish{}

\DeclareMathOperator*{\argmax}{arg\,max}

\newcommand{\dsone}{\textsc{Tune}\xspace}
\newcommand{\dstwo}{\textsc{Big}\xspace}
\newcommand{\dsthree}{\textsc{Chrono}\xspace}
\newcommand{\dstwotest}{\textsc{SelfPack}\xspace}
\newcommand{\extendset}{\textsc{Big+}}
\newcommand{\virustotal}{VirusTotal\xspace}
\newcommand{\peframe}{PEframe\xspace}
\newcommand{\manalyze}{Manalyze\xspace}
\newcommand{\peid}{PEiD\xspace}
\newcommand{\die}{DIE\xspace}
\newcommand{\cisco}{Cisco\xspace}
\newcommand{\knn}{\textsc{KNN}\xspace}
\newcommand{\nbc}{\textsc{NBC}\xspace}
\newcommand{\gnbc}{\textsc{GNBC}\xspace}
\newcommand{\bnbc}{\textsc{BNBC}\xspace}
\newcommand{\lin}{\textsc{Lin}\xspace}
\newcommand{\linreg}{\textsc{LR}\xspace}
\newcommand{\linsvm}{\textsc{LSVM}\xspace}
\newcommand{\dt}{\textsc{DT}\xspace}
\newcommand{\dl}{\textsc{DL}8.5\xspace}
\newcommand{\rf}{\textsc{RF}\xspace}
\newcommand{\gbdt}{\textsc{GBDT}\xspace}
\newcommand{\svm}{\textsc{KSVM}\xspace}
\newcommand{\nn}{\textsc{NN}\xspace}
\newcommand{\mlp}{\textsc{MLP}\xspace}

\begin{document}
\begin{frontmatter}

\title{Analysis of Machine Learning Approaches to Packing Detection}
\author[add1]{Charles-Henry Bertrand Van Ouytsel\corref{cor1}}
\ead{charles-henry.bertrand@uclouvain.be}

\author[add1]{Thomas Given-Wilson\corref{}}
\author[]{Jeremy Minet\corref{}}
\author[]{Julian Roussieau\corref{}}
\author[add1]{Axel Legay\corref{}}
\address[add1]{INGI, ICTEAM, Universite Catholique de Louvain, Place Sainte Barbe 2, LG05.02,01, 1348 Louvain-La-Neuve, Belgium}
\cortext[cor1]{Corresponding author}

\date{May 2, 2021}

\begin{abstract}
Packing is an obfuscation technique widely used by malware to hide the content and behavior of a program.
Much prior research has explored how to detect whether a program is packed.
This research includes a broad variety of approaches such as entropy analysis, syntactic signatures and more recently machine learning classifiers using various features.
However, no robust results have indicated which algorithms perform best, or which features are most significant.
This is complicated by considering how to evaluate the results since accuracy, cost, generalization capabilities, and other measures are all reasonable.
This work explores eleven different machine learning approaches using 119 features to understand:
which features are most significant for packing detection;
which algorithms offer the best performance; and
which algorithms are most economical.
\end{abstract}
\end{frontmatter}

\section{Introduction}
\label{sec:introduction}

Malware detection represents a significant and expensive problem for current computer security.
Since the vast majority of malware detection programs are based on signatures, this means that they are reactive in nature and so the malware writers are typically one step ahead.Further, new malware are constantly being created, the AV-Test Institute \cite{AVTest} registers an average 350,000 new \emph{malware} (malicious programs) every day.

Malware analysis techniques used to identify, understand, and detect malware are typically divided into two approaches: \emph{static analysis} and \emph{dynamic analysis}. Static analysis operates by examining the sample program without executing the program. This can range from very simple analysis of strings or bytes or the code, to complex disassembly and reconstruction of key program behaviours and features. Most malware detection programs use static analysis on simple features such as strings. Dynamic analysis operates by executing the sample program in a protected environment, e.g.~a sandbox, and observing the program's behavior such as recording API calls, network activity, process creation, etc. While dynamic analysis allows more in-depth inspection, the cost of starting and running a sandbox is significant. Most detection engines cannot afford to perform dynamic analysis, and so dynamic analysis is usually only done for signature creation or by malware analysis teams to understand new samples.

\emph{Packing} is a widely used technique strategy that is able to evade or disrupt many malware analysis techniques.
Packing operates using various methods such as: compressing or encrypting sections of the program; encoding the program in a virtual machine; or fragmenting program behavior.
All of these make both static and dynamic analysis much more difficult, since they obfuscate the static features and program behavior, and also make the dynamic execution behavior more complex and harder to observe.
Packing techniques are typically implemented with a small stub section of the program that performs the initial decompression/decryption or runs the virtual machine and thus allows the program to execute its behavior.
While packing is used by benign programs (typically for compression or to protect intellectual property) WildList~\cite{Wildlist} states that 92\% of packed programs hide harmful behavior.

\begin{figure}[!ht]
\centering
  \includegraphics[width=0.5\linewidth]{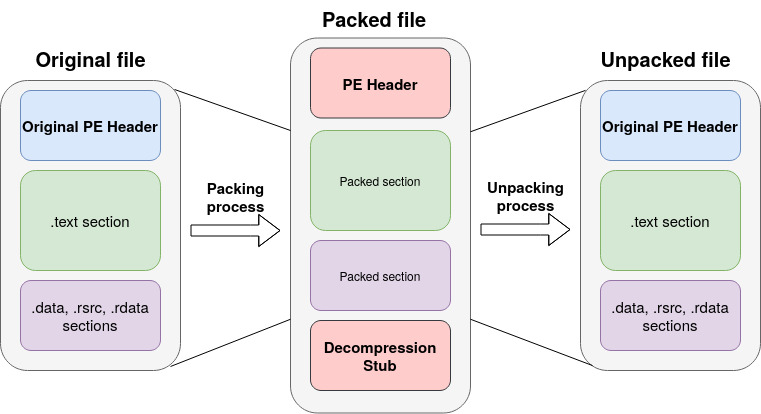}
  \caption{Life-cycle of a packed program}
  \label{fig:packing_process}
\end{figure}

Figure \ref{fig:packing_process} illustrates the life-cycle of a typical packed program.
The original program is compressed and the compressed program is stored in the ``Packed section''. Another section is also added containing the decompression stub routine.
When the program executed, the decompression stub decompresses the packed section and then executes the decompressed program. .

To create an effective malware detection or classification engine, it is critical to be able to detect whether a program being analysed is packed.
A packed program can be unpacked or treated with greater suspicion to improve the analysis and response.
This approach was developed by Perdisci et al. \cite{perdisci_classification_2008} to efficiently distinguish if a program is packed and then deliver the packed programs to a universal unpacker (a program to reverse the packing).
Combined with a fast and efficient static malware detector such as the one proposed by Baldangombo et al. \cite{baldangombo_static_2013}, a complete malware detection engine can offer high-end performance malware detection and classification. %

Packing detection and classification have been widely studied in the literature 
\cite{DBLP:journals/ieeesp/LydaH07,DBLP:journals/ieeesp/LydaH07,choi2008pe,jeong2010generic,raphel2015information,ugarte2011structural,ugarte2012countering,ugarte2014adoption,perdisci_classification_2008,choi2008pe,jeong2010generic,ugarte2011structural,ugarte2014adoption,raphel2015information,zakeri2015static,margaria_tutorial_2018,DBLP:conf/codaspy/AhmadiUSTG16,DBLP:journals/corr/abs-1802-10172,biondi2019effective}.
These works consider packing detection through a wide variety of approaches to detection and classification such as using rules, heuristics, or machine learning.
There is also a wide variety of potential static (and dynamic) features considered for which may be most significant for accurate (and efficient) packer detection and classification.
Static features are preferred as the basis for detection and classification since they can be easily extracted with low cost, and thus are widely used as the basis for popular antivirus such as ClamAV~\cite{ClamAV} or tools such as PEiD~\cite{peid}.

This work addresses the challenges with the current state-of-the-art on packing detection that various ML approaches on various features appear effective, but there is a lack of clarity on which ML techniques and features are the most effective.
To address this challenge, this work explores 11 ML algorithms over a variety of data sets with various forms of ground truth, samples, and to measure different kinds of performance.
In particular this works considers 11 different ML algorithms that have been used for packing detection.
The ground truth generation for data sets is performed using 5 different labeling approaches and a majority vote among them \cite{biondi2019effective}.
The features are also processed in different forms since data processing is a significant factor for the effectiveness of ML algorithms.
The 11 ML algorithms are then each tuned according to their own (hyper) parameters to gather information on which parameters are most significant to each algorithm.
The 11 tuned ML algorithms are then used to explore which features of 119 static features are the most significant for packer detection.
The 11 ML algorithms are then applied to larger data sets to explore their effectiveness considering different metrics. Their resilience against different known packers is also assessed by packing malware ourself.
Finally, an economical analysis (how accuracy degrades over time and on new samples) is considered to explore the effectiveness of the 11 ML algorithms over time and as an tool.

\medskip

\label{RQs}

The challenges explored in this work can be focused into the following research questions:
\begin{itemize}

\item [RQ1] \textbf{Which features are most significant for packing detection?} We explore relative relevance of features present in literature into the classification decision process.

\item [RQ2] \textbf{Which ML classifiers are effective for packing detection?} We study performances of our classifiers over large dataset and against specific packers (present or not in the dataset) to discover their strengths and weakness.

\item [RQ3] \textbf{Which ML algorithms perform well in long term for packing detection?} We train our classifiers on a huge dataset and evaluate them over different samples recorded later. Observing evolution of their effectiveness reveals us how accuracies of classifiers change over time and which classifiers are more economical in term of training.
\end{itemize}

\medskip

\subsection{Related Work}
\label{ssec:related}

This section briefly overviews popular approaches to packing detection.

Syntactic signatures are a popular approach to packer detection that can be extremely cheap to compute and still achieve high accuracy for known packers.
Signatures such as those used by YARA~\cite{YARA} typically consider simple static features such as byte sequences, strings, and file features such as size or hash.
Although such signatures are effective for exactly matching known samples, they perform less effectively against even minor permutations and can be easily evaded \cite{margaria_tutorial_2018}.

Entropy metrics have been shown effective in detecting packing since packed files often have much higher entropy than unpacked files. Such approaches were considered in \cite{DBLP:journals/ieeesp/LydaH07,choi2008pe,jeong2010generic,raphel2015information,ugarte2011structural,ugarte2012countering,ugarte2014adoption}. Lyda et al.~\cite{DBLP:journals/ieeesp/LydaH07} were first to address the problem of packing detection using entropy in an academic publication. Their approach was focused on global entropy (computed over the whole program) used as a proxy to detect packing. Their dataset was cleanware for non-packed samples while packed samples where obtained by employing well-known packers on each cleanware sample. Although not verified experimentally, their approach was calculated to target a 3.8\% false positive rate and 0.5\% false negative rate. The concept was then improved by evaluating entropy for each section of the sample in the paper of Han and Lee~\cite{han2009packed}. They use 200 cleanware for non-packed samples and 200 malware from a honeypot for packed samples, obtaining a detection rate of 97.5\% with false positive of 2.5\%. However, methods focusing exclusively on entropy are vulnerable to adversaries. For example, inserting selected set of bytes in an executable in order to keep the entropy of the file low such as in the Zeus malware family \cite{ugarte2012countering}.

\emph{Machine learning} (ML) techniques using supervised learning on variety of features extracted statically have been investigated in different works \cite{DBLP:journals/ieeesp/LydaH07,choi2008pe,jeong2010generic,ugarte2011structural,ugarte2014adoption,raphel2015information,zakeri2015static, DBLP:conf/codaspy/AhmadiUSTG16,DBLP:journals/corr/abs-1802-10172,biondi2019effective}.
Perdisci et al.~investigated different ML methods in packing detection : Naive Bayes Classifier, decision tree, ensemble of decision trees, k-nearest neighbors and Multi-layer perceptrons. They consider 9 features (also considered in this work) related to section names and properties, import access table (IAT) entries and several entropy metrics in~\cite{perdisci_classification_2008} while using a dataset of 5,498 PE files (benign and malicious PE). Multi-layer perceptrons performed well in their work (98.91\% test accuracy) compared to other investigated methods. That is why, multi-layer perceptron algorithm is also investigated in our work. We will go deeper by presenting more insights on features relevance, tuning and performing an economical analysis. Our bigger data set will additionally allow more meaningful measure of accuracies. In~\cite{perdisci2008mcboost}, they add N-grams features to their work and include the packing detection process into a workflow to analyse and detect malwares.

Diverse new features were proposed by malware researchers, for instance in \cite{arora2013heuristics,santos2011collective,ugarte2014adoption} like structural PE attributes, heuristic values, entropy-based metrics or different ratio (raw data per virtual size ratio or the ratio of the sections with virtual size higher than raw data for instance). Although more expensive to extract, some dynamic features have also been studied in the literature, such as using evolution of the import address table (IAT) table during the unpacking process~\cite{DBLP:conf/ccs/BonfanteFMRST15} or basing analysis on `write-then-execute' instructions typical in packed samples~\cite{DBLP:conf/ccs/Cheng0FPCZM18,kang2007renovo}.

Recent work~\cite{biondi2019effective} has surveyed the literature, explored the feasibility of using machine learning techniques for packing detection and produced a list of 119 commonly used features (the same used in this work). Most of this work focus on defining the methodology and on developing efficient techniques to extract interesting features statically.
In their paper, Biondi et al.~divide them into six categories: Byte entropy, Entry bytes, Import functions, Metadata, Resource and Sections. Moreover, they focus on impact of each categories. Our work investigates this set of 119 features (See \ref{sec:appendix_extracted_features}) by considering all of them as independent. Their paper also discuss few machine learning algorithms based on those features to show the feasibility of the approach (i.e. Bayes classifier, Decision trees and Random Forest).They studied these algorithms over a dataset of 280,000 samples in terms of effectiveness, robustness and efficiency (related to computational cost). They conclude their study by showing that accepting a small decrease in effectiveness can increase efficiency by many
orders of magnitude. We expand their set of studied ML classifiers and apply similar analysis on their performance.  

The problem of packing detection has also been extended to packing classification~\cite{sun2010pattern}. Different classifier have been compared in this perspective using pattern recognition techniques on randomness profiles of packers. In \cite{bat2013dynamic,bat2017entropy,jeong2010generic}, entropy patterns extracted dynamically (i.e., by executing the binary) are considered as an efficient way to classify packers. Knowing the file is packed, the entropy of the whole file is recorded after each JMP instruction until entropy ceases to change. In~\cite{bat2017packer}, the authors use similar techniques to determine if multiple packers have been used on the same program. The problem of \emph{known} packer classification has been demonstrated as a trivial task in~\cite{aghakhani2020malware}. Using samples from each packer (class), their classifier achieved precision and recall
greater than 99.99\% for each class. However, the authors have highlighted challenges related to problems of packing detection.

Unsupervised learning has recently also been investigated~\cite{noureddine2021se} to build clusters related to packers and identify variants of known packers or new packers. However, this is beyond the scope of this work, as the focus here is on exploring supervised learning algorithms.

\medskip

The structure of the paper is presented hereafter.
Section~\ref{sec:back-ML} recalls background on the machine learning algorithms used in this work.
Section~\ref{sec:Exp_setup} presents the construction of datasets used in this work, the methodology applied to answer our research questions and configurations used for our ML algorithms (i.e.: pre-processing and hyper-parameter tuning).
Section~\ref{sec:FeaturesAnalysis} explores the relevance of different feature in the decision process of classifiers, including feature selection methods and principal component analysis.
Section~\ref{sec:Effectiveness} evaluates the effectiveness of the ML algorithms on a large data set.
Section~\ref{sec:Economics} explores the economics of the trained packing detectors over a chronological data set.
Section~\ref{sec:validity} discusses the validity of the results and some observations on dynamic features.
Section~\ref{sec:Conclusion} concludes.

\section{Background on Machine Learning algorithm}
\label{sec:back-ML}
This section recalls useful background information about eleven machine learning algorithm used in this paper.

\subsection{K-Nearest Neighbors (\knn)}
\label{sssec:knn}

The \emph{K-Nearest Neighbors} (\knn) ML algorithm considers each input data as a point in a feature space with a particular label. When given a new data point to classify \knn will find the K nearest points in the feature space in terms of  a defined distance (e.g.~Euclidean distance) and classify the new data point as a plurality vote of its neighbors within the defined distance. In the case of $k=1$, the input is simply assigned the label of its nearest neighbor.

\begin{figure}[!ht]
\centering
  \includegraphics[width=0.40\linewidth]{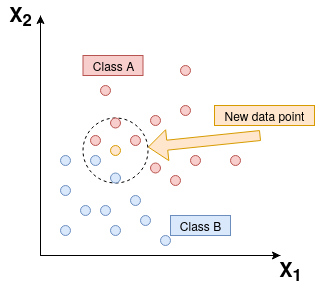}
  \caption{Binary classification involving two features ($X_1$ and $X_2$) and using a 5-Nearest Neighbors classifier}
  \label{fig:KNN}
\end{figure}

An example is shown in Figure~\ref{fig:KNN} with two labels (red and blue) where the new data point (orange) will be assigned the color red according to the majority vote. This classification algorithm requires complete information about the whole training data set to classify each new data point and thus requires significant memory use in practice for larger data sets. 

\subsection{Naive Bayes Classifiers (\gnbc \& \bnbc)}
\label{sssec:nbc}

\emph{Naive Bayes classifiers} (\nbc) are probabilistic models based on Bayes theorem:
\begin{equation}
\label{eqn:Bayes}
\Pr(C_k|x_i)=\frac{\Pr(x_i|C_k)\Pr(C_k)}{\Pr(x_i|C_k)\Pr(C_k)+\Pr(x_i|\neg C_k)\Pr(\neg C_k)} \; 
\end{equation}
where
$C_{k}$ is a class in the set of classes $k\in\{1,..,K\}$ and
$x_i$ is a feature in the set of feature vectors $i \in \{1,..,n\}$ and
$\Pr(a)$ is the probability of $a$ (where $a$ could be a class label or feature vector) and
$\Pr(a|b)$ is the probability of $a$ given $b$ (where $a$ is a class label and $b$ is a feature vector or vice versa).
These classification use Equation \ref{eqn:Bayes} to evaluate probability to observe label $C_k$ for a sample according to value of feature vector $x_j$. These estimated probabilities are then combined to construct the classifier as in Equation~\ref{eqn:BayesClass} where $\hat{y}$ is the estimated label of the sample:

\begin{equation}
\label{eqn:BayesClass}
\hat{y} = \argmax_{k\in \{1,..,K\}} p(C_k) \displaystyle \prod_{i=1}^{n} p(x_i|C_k)
\end{equation} 
The \emph{naive} adjective of the classifier comes from the assumption that each data point's feature is independent from each other feature. Although this assumption does not usually hold in general \nbc is known to work well in practice.

Two variant of the \nbc are investigated in this paper: \emph{Gaussian} (\gnbc) based on normal distribution and \emph{Bernouilli} (\bnbc) where features are considered as independent booleans.
Although \nbc are among the simplest ML algorithms, they offer good accuracy in some areas and have low performance costs.

\subsubsection{Linear Models (\linreg \& \linsvm)}
\label{sssec:lin}

In \emph{linear models} (\lin) ML algorithms a model is built from training data as a linear combination of features. The model has the form given in Equation~\ref{eq:linEQ} below:
\begin{equation}
\label{eq:linEQ}
y = x[1] \cdot w[1] + x[2] \cdot w[2] + ... + x[p] \cdot w[p] + b
\end{equation}
where
$y$ is the prediction calculated from
$x[i]$ is each feature indexed by $i$ in the feature vector $x$ and
$w[i]$ is the learned weight for each $i$ in the length of the feature vector and
$b$ is the offset.

Since our focus is binary classification, the predicted value is set to a threshold at zero.
Thus, a $y$ value greater than zero corresponds to label $0$ while a value less than zero corresponds to label $1$.
Therefore, the decision boundary can be represented as a line or a plane that separates the data in two classes.
Since \lin are a very large family of algorithms this paper considers two forms of \lin: logistic regression and linear support vector machines. 
\begin{figure}[!ht]
\begin{minipage}[b]{0.50\linewidth}
\includegraphics[width=0.8\linewidth]{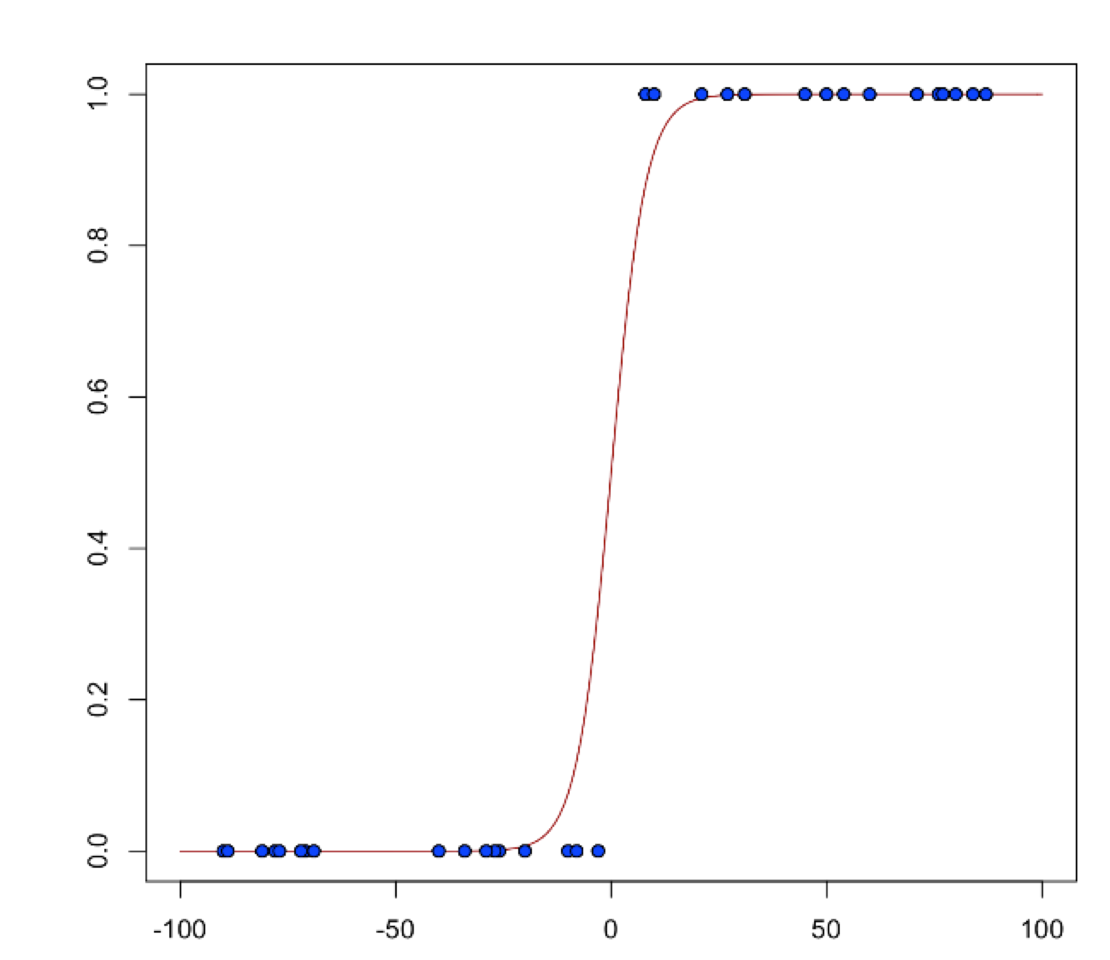}
\caption{Logistic regression \cite{logreg1}}
\label{fig:LM1}
\end{minipage}
\begin{minipage}[b]{0.50\linewidth}
    \includegraphics[width=0.8\linewidth]{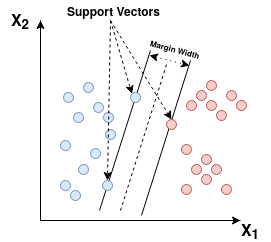}
\caption{Linear support vector machine.}
\label{fig:LM2}
\end{minipage}
\end{figure}

\emph{Logistic regression} (\linreg) is based on the logistic function: 
\begin{equation}
\label{eq:linreg}
\sigma(z)=\frac{1}{1+e^{-z}}\; .
\end{equation}
that is also illustrated by the red curve in Figure~\ref{fig:LM1}.
To obtain optimal weights $w$ for \linreg, maximum likelihood estimation is used.
This involves finding weights $w[i]$ that maximize the log-likelihood function defined as follows:
\begin{equation*}
L(w) = \displaystyle \prod_{i=1}^{n} \sigma(w^Tx_(i))^{y_i} \cdot  [1 - \sigma(w^Tx_(i))]^{1-y_i}\; .
\end{equation*}

A \emph{Linear support vector machine} (\linsvm) is based on parallel hyperplanes that separate the classes of data, so that the distance (or margin) between them is maximised.
Samples used to construct the two planes 
are called \emph{support vectors}.
The equation used to define these is as follows:
\begin{equation}
\label{eq:planesSVM}
w^Tx-b=1 \quad\textrm{ and }\quad w^Tx-b=-1\; .
\end{equation}
Figure~\ref{fig:LM2} shows the two planes (here two lines), support vectors used to build the planes, and the margin width that we try to maximize here.

\subsection{Tree-based Models (\dt, \gbdt, \rf, \& \dl)}
\label{sssec:dt}

\begin{figure}[!ht]
\centering
  \includegraphics[width=0.35\linewidth]{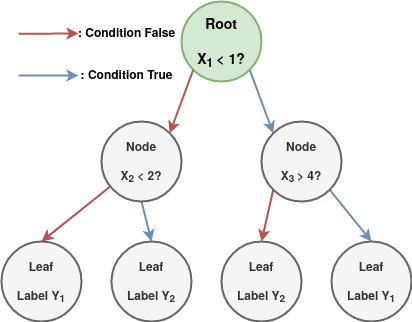}
  \caption{Example of a simple decision tree}
  \label{fig:DT}
\end{figure}

\emph{Decision trees} (\dt) are classification models based on the sequential application of simple binary rules. They are structured as a directed tree as illustrated in Figure~\ref{fig:DT}. Starting at the root (in green), the tree is traversed towards a leaf (containing a label) by selecting an edge at each node according to a simple rule (e.g.~in Figure~\ref{fig:DT}, the rule at the root is ``if $X_1$ of the sample is smaller than 1 go to right branch, else go to left branch'').
In the building of a decision tree, all features are considered. Different combinations of splits are considered and the best one is selected according to a cost function (such as gini impurity or entropy). The choice of effective split being of first importance in term of tree size and cost, although features relevance and features selection is an important concern which can assist with pruning or depth limitation.

Simple \dt models can be combined to construct more complex classifiers solving variance and bias that single \dt can suffer from.
The concept is to combine many weak learners to shape a stronger one at the cost of more resource consumption.
Two of these classifiers, known to have demonstrated good performance on a diversity of data sets are considered in this work: Random Forests and Gradient Boosted Decision Trees.

\emph{Random Forests} (\rf) represent the combination of slightly different \dt. While a single \dt may be very effective there is a risk be overfitting the training data. \rf therefore address this risk by considering outcomes from several \dt, resulting in a reduced overfitting while maintaining the predictive power of each tree. The random nature of this algorithm resides in the creation of the \dt, which are built upon a randomly generated set of samples (this process is called \emph{bagging}). Each tree is then trained individually and their outputs are aggregated with each other in a majority vote.

\emph{Gradient Boosted Decision Trees} (\gbdt) is another approach to address overfitting that works by using a technique called \emph{boosting} (while \rf uses bagging). In this process, each tree is constructed sequentially with each subsequent tree trying to correct the errors of the previous trees. Typically \gbdt start with extremely shallow trees, often called weak learners, which provide good prediction only on some parts of the data. Increasingly trees are then grown iteratively to increase the global performance by reducing the loss margin. In comparison to \rf, \gbdt produces shallower trees. Hence, the model is more economical in terms of memory but harder to tune.

In addition to the traditional approach, a recent version of \dt is also considered in this work, namely \emph{\dl} \cite{nijssen2020learning}, working only with binary data (implementation available at \cite{DL8.5}).
\dl differs from classical approaches by using a cache of item sets combined with branch-and-bound search in order to create less complex trees and avoid unnecessary computations.

\subsection{Kernelized Support Vector Machines (\svm)}
\label{sssec:ksvm}

\begin{figure}[t]
\begin{minipage}[b]{0.5\linewidth}
    \includegraphics[width=0.8\linewidth]{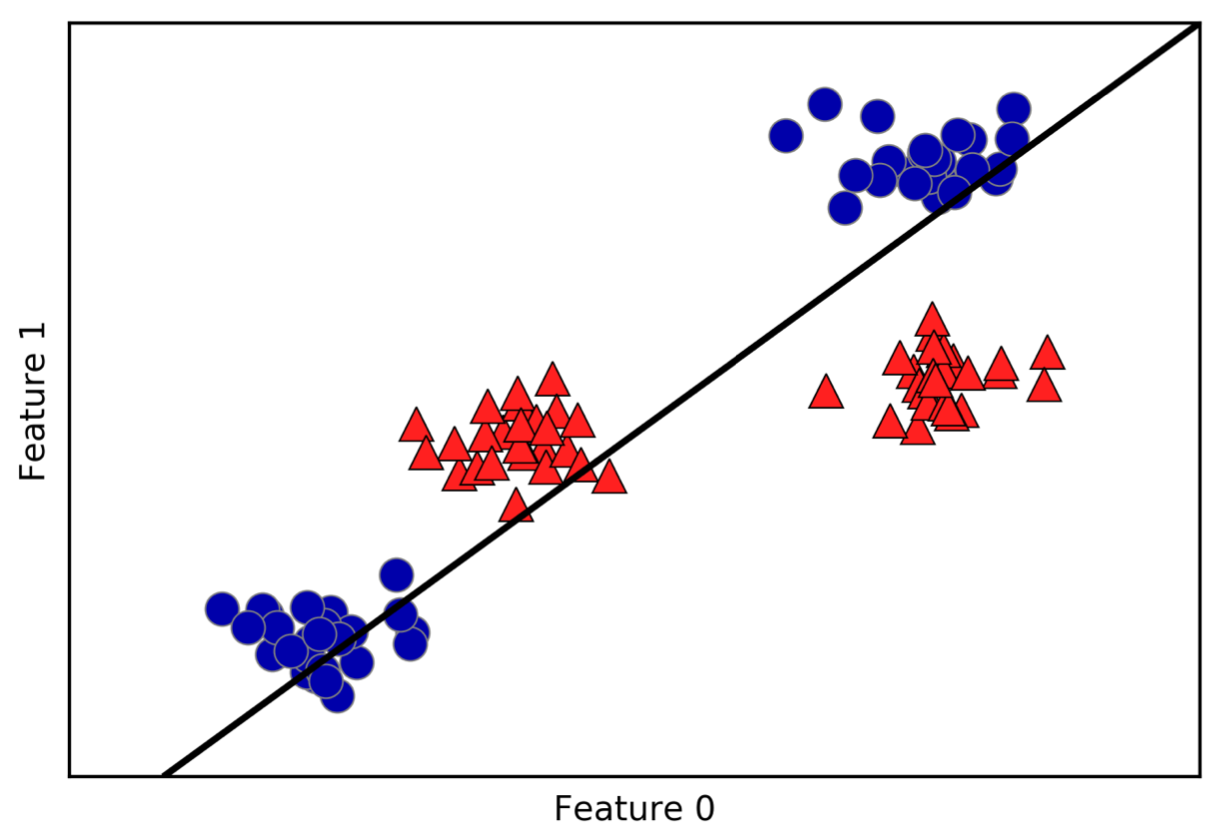}
    \caption{Decision boundary found by \lin \cite{ML}.}
    \label{fig:SVM1}
\end{minipage}
\begin{minipage}[b]{0.5\linewidth}
\includegraphics[width=0.8\linewidth]{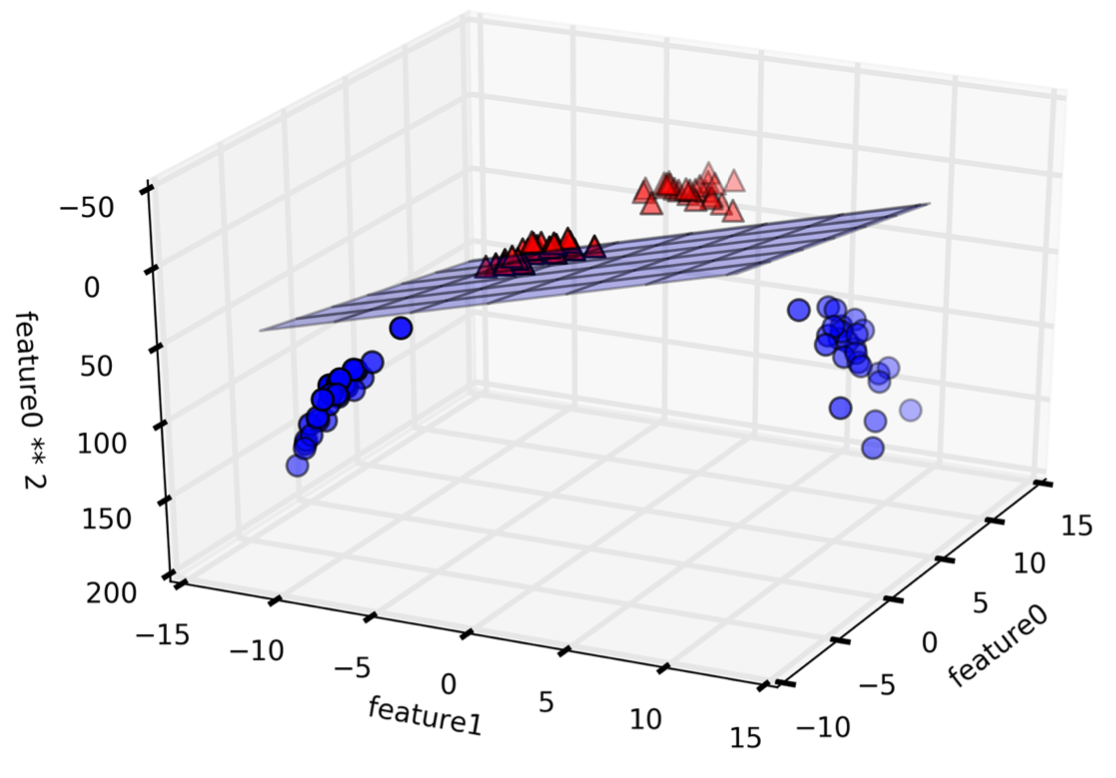}
\caption{Decision boundary found by \svm \cite{ML}.}
\label{fig:SVM2}
\end{minipage}
\end{figure}

\emph{Kernelized Support Vector Machines} (\svm) are an extension of \lin which allows separation of data not linearly separable by using hyper-planes.
In \svm low dimensional data are converted to higher dimensional data using a \emph{kernal function} (e.g.~polynomial, Gaussian, Sigmoid, etc.) and the planes used to separate the data are now higher dimensional hyperplanes.
For example, on Figure~\ref{fig:SVM1} is shown how \lin would try to find planes to classify the red and blue data points.
On Figure~\ref{fig:SVM2} the data points are converted to a higher dimension and a higher dimensional hyperplane is used to separate the points, leading to a much more effective separation.

Although \svm work well on a variety of datasets, they are known to suffer from performance degradation when scaled to a large number of samples.

\subsection{Neural Networks (\mlp)}
\label{sssec:nn}

\begin{figure}[!ht]
\centering
  \includegraphics[width=0.50\linewidth]{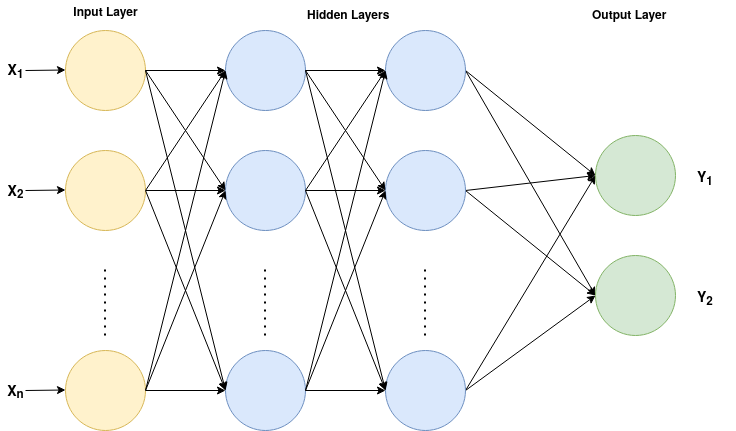}
  \caption{Multi-Layer Perceptron with two hidden layers.}
  \label{fig:MLP}
\end{figure}

\emph{Neural Networks} are mainly used in deep learning and have been employed in thousands of different classifiers for specific purposes.
This work uses  
the most popular variation of \nn: \emph{Multi-Layer Perceptrons} (\mlp) algorithm.
\mlp can be viewed as a generalization of \lin that is constituted with various layers of neurons. Each neuron is a unit taking its input applying a weighted sum of its features and passing it through a non-linear function before forwarding it to the subsequent layer as illustrated in Figure~\ref{fig:MLP}.

Each neuron of each layer with input features $X$ will apply the following equation to calculate its output $y$:
\begin{equation}
y = f(x[0] \cdot w[0] + w[1] \cdot x[1] + ... + w[p] \cdot x[p] + b)
\end{equation}
where $f$ is a non-linear function (e.g.~logistic function like Equation~\ref{eq:linreg} or hyperbolic tangent) applied to a weighted sum on $x$ features.
With \mlp, this process of weighted sums is repeated multiple times. Data are supplied to the input layer, then there may be one or more hidden layers representing intermediate processing steps, and eventually predictions are made on the output layer.

The main advantage of using \nn is that they have demonstrated their efficiency to deal with large data sets and come up with incredibly complex models at the cost of training time, wise parameter tuning and accurate pre-processing.

\section{Experimental Setup}
\label{sec:Exp_setup}

This section details the experimental setup used in this work. 
This is divided into three areas:
construction of data sets;
experimental methodologies; and
pre-processing and hyper-parameter tuning of the ML algorithms.

\subsection{Data Sets and Labeling Tools}
\label{ssec:sets}

All data sets are built from a feed of malware provided by Cisco that offers 1,000 new (assumed) malware samples per day \cite{Shadow}.
The construction of data sets begins by taking samples of malware from the feed that are collected over a given time period. For each sample, 119 static features are (attempted to be) extracted\footnote{Occasional errors in the extraction software caused a few samples to be discarded, although this is an insignificant number and so ignored for the rest of this work.}. These features are a collection of all those used in other works and identified in the work of Biondi et al.~\cite{biondi2019effective}.

Once gathered, these samples need to be labeled as packed or not packed to obtain a ground truth for training ML classifiers. In this work a similar approach to the one of \virustotal was taken; to employ multiple \emph{detector} engines for classification and then build on these results. There were five classification engines used, described below.
\begin{itemize}
\item \textbf{\peframe}~\cite{peframe} This is an open source tool written in Python and freely available to perform static analysis of portable executable (PE) files, object code, DLLs and others. Its utilization will be restricted to packer detection but the tool can also be used to obtain other digital forensics information such as macros, anti-forensics techniques, etc. 
\item \textbf{\manalyze}~\cite{manalyze} This is a static analyzer mainly used by large enterprises and composed of several plugins. Its packer detection plugin is based on signatures and custom heuristics. Signatures are generally related to names of PE sections (e.g.~UPX renames some of the section after packing as UPX0, UPX1, etc.) while heuristics are more related to anomalies in the PE structure (unusual section names, sections both writable and executable, etc.) or entropy-based features. 
\item \textbf{\peid}~\cite{peid} This is a widely known detector, freely available and signature based. Its signatures only contains low-level byte patterns which can match either at the entry point or more generally anywhere within a PE file.  
\item \textbf{\die} Detect-it-Easy ~\cite{die} is an open source architecture of signatures written in javascript which allows more complex and fine tuned signatures. 
\item \textbf{\cisco} This was an in-house engine that Cisco used on some samples. This is not publicly available and was only used for some of experiments due to limited availability.
\end{itemize}

A sample is considered as packed if a majority of detectors label the sample as packed, in this case three out of five detectors. When the Cisco detector was unavailable, a threshold of two out of four was used.

\bigbreak

A first data set \dsone of 43,092 samples is used for parameters tuning and finding optimal representation of the dataset (by feature selection and Boolean conversion).
This data set consists of samples collected from 2019-06-15 to 2019-07-28 and according to the labeling has 10,009 packed samples and 33,083 not packed samples.
Labels for all five label engines were available for this data set.

\bigbreak

A second much larger data set \dstwo of 95,876 samples is used for the broader experimental results in this work.
This data set consists of samples collected from 2019-10-01 to 2020-02-28 and according to the labeling has 23,894 packed samples and 71,982 not packed samples.
Since labels from the \cisco engine were exclusively available for the samples in \dsone, this data set was generated to carry out larger experiments on the machine learning and features, but had reduced label information.
An extended data set denoted \extendset$_{f}$ is the data set \dstwo with 500 samples taken at random from \dsone and packed with the packer $f$ added to \dstwo.
The label information is the same for samples originally from \dstwo, the extended samples have known packer.

Specific data sets \dstwotest$_{f}$ of approximately 12,036 samples are samples collected from 2020-06-15 to 2020-06-30 and packed ourself with known packer $f$ .

\bigbreak

A third data set \dsthree of 37,794 samples is used for the economical analysis results in this work.
This data set consists of samples collected from 2020-04-01 to 2020-05-26 and according to the labeling has 8745 packed samples and 29,049 not packed samples. This data set has also reduced label information as \dstwo.

\subsection{Methodology}
\label{ssec:Methodology}

To address the research questions effectively it is necessary to properly pre-process the data sets and tune the ML algorithms appropriately.
This is detailed in Section~\ref{ssec:Preprocessing}, however an overview is as follows.
The pre-processing includes Boolean conversion of non-Boolean data, bucketing, and normalization of feature values.
Once the data sets have been pre-processed, the hyper-parameters of all the ML algorithms are tuned using \dsone to find the best configurations for later use.

The methodology used to address the various research questions in this paper is split into several parts based upon the research question being considered.
\begin{itemize}
\item
\textbf{RQ1} is addressed by exploring which features are the more significant in different ML algorithms. Due to there being many different ML algorithms several approaches are used for reducing the number of features including: feature coefficients, K-best features, and principle components analysis. These approaches are applied to all 11 ML algorithms using \dsone and experimentally evaluated with $10$-fold cross-validation and $90\%$ of the data for training and $10\%$ for testing. Details on the approach and results are presented in Section~\ref{sec:FeaturesAnalysis}.

\item
\textbf{RQ2} is addressed by taking all the ML algorithms along with all the lessons learned on data processing, ML tuning, and feature selection to evaluate their overall effectiveness on the larger data set \dstwo.
The analysis is performed by using $10$-fold cross-validation on \dstwo and taking $90\%$ of the data for training and $10\%$ for testing.
Further, the adaptability of the 11 ML algorithms is evaluated by comparing the accuracy when the data set is extended with more samples of a specific packer.
This evaluates the adaptability of the algorithms as their training data changes, and to evolutions in packer families.
Details of the results and experiments are presented in Section~\ref{sec:Effectiveness}.

\item
\textbf{RQ3} is addressed by training the ML algorithms (developed and tuned as above) on \dstwo and then evaluating them on \dsthree.
Observe that \dsthree is chronologically after \dstwo and thus indicates a representation of the evolution of malware over time (there is a time gap of a month between last malware of \dstwo and first malware of \dsthree to focus on packed samples which would be release later than the training set).
To further assess the economic and chronological performance of the ML algorithms, the testing on \dsthree is considered over smaller chronological periods (four periods are considered, each one consisting of two weeks).
This evaluates the effectiveness of the various ML algorithms over time and allows to calculate the effective cost of retraining to maintain a desired level of accuracy.
Note that since the experiments here consider using data sets from various times, the whole early data set is used for training and not cross-validation is applied. Details of the approach and results are presented in Section~\ref{sec:Economics}.
\end{itemize}

\medskip

All the ML algorithm implementations used are from the scikit-learn library~\cite{scikit-learn} version 0.23.2. except the \dl algorithm~\cite{nijssen2020learning} where the implementation on github is used (that implement the same functions as the scikit interface).
All experiments here are performed on a desktop PC with an Intel Core i7-8665U CPU (1.90GHz x 8) and 16GB RAM running Ubuntu 18.04.5.

\subsection{Pre-processing and Hyper-Parameter Tuning}
\label{ssec:Preprocessing}

This section overviews the data pre-processing and ML algorithm hyper-parameter tuning used in the later experiments.
Pre-processing includes the consideration and application of Boolean conversion, standardization and normalization of the data.
For each algorithm, a grid search is applied to find the best combination of pre-processing and hyper parameters.

\subsubsection{Data Pre-processing}
\label{sssec:pre_processing}

\begin{figure}[!ht]
\begin{minipage}{0.45\textwidth}
\begin{itemize}
\item $42\%$ have value 1024 (in blue).
\item $34\%$ have value 4096 (in orange).
\item $16\%$ have value 512 (in green).
\item $7\%$ have value 1536 (in red).
\item $1\%$ have other values (in yellow).
\end{itemize}
\end{minipage}
\hfill
\begin{minipage}{0.5\textwidth}
\centering
\includegraphics[width=\linewidth]{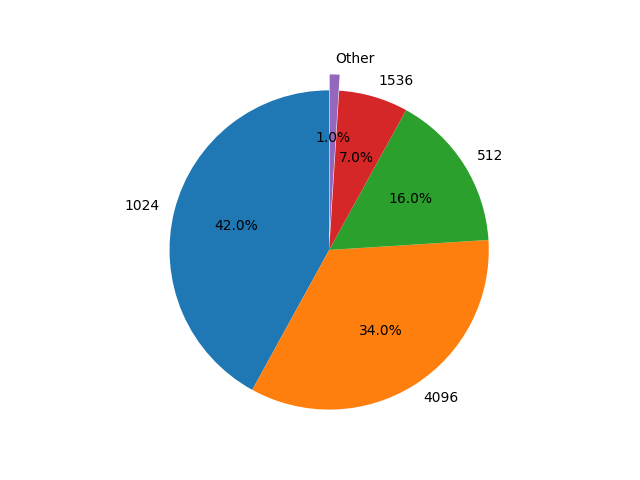}
\end{minipage}
\captionof{figure}{Distribution of values for feature 16 (total header size in bytes).}
\label{fig:fv_repartition}
\end{figure}

Of the 119 static features considered for each sample, only 16 of them are Boolean values.
Some of the ML algorithms such as \dl requires Boolean features (for details see Section~\ref{sec:back-ML}).
Therefore it is necessary to convert non-Boolean values to Boolean values for use in these algorithms. 

This conversion begins by observing the value distribution of \dsone features. For example, Figure~\ref{fig:fv_repartition} shows the distribution of feature 16; namely header size in bytes.
Bucketing is then applied which gathers similar values into buckets and hence reduces the range of possible values.
For example: value 1 is mapped to 1024, 2 is mapped to 4096, 3 is mapped to 512, 4 is mapped to 1536 and 0 is mapped to other possible values.
Then one-hot encoding is used to convert these bucket values into Boolean values. Therefore in the case of the feature 16, five Boolean features will be used in the new dataset to represent header size. This method induces an important increase in term of the number of features and it is therefore important to efficiently limit the number of buckets for each feature.
Details on the results of this pre-processing via bucketing and conversion to Boolean values is given in \ref{sec:appendix_boolean_conversion}. 

Initially Boolean conversion is used only in the \bnbc and \dl ML algorithms which requires Boolean data.
However, since other algorithms could benefit from this conversion, Boolean features conversion is also tested on all the ML algorithms to observe the impact.
Final choice for each algorithm is investigated in Section~\ref{sssec:hyper_tuning}. 

\medskip

To work most efficiently, some ML algorithms require their data to be \emph{normalized} (values $\in [0;1]$) or \emph{standardized} (considered as a Gaussian distribution $\in [-1;1]$). 
For example, \gnbc requires normalization to produce an efficient and coherent model.
Like for the Boolean conversion, although many ML algorithms do not require normalization or standardization, they may be able to benefit from this data pre-processing.
This is investigated below.

\subsubsection{Hyper-Parameters Tuning}
\label{sssec:hyper_tuning}

Each of the 11 ML algorithms has their own hyper-parameters that can be tuned to improve their performance on different kinds of data sets and classification problems.
In this section a grid search is applied to each ML algorithm to obtain the best combination of pre-processing and hyper-parameters.
The best combination is chosen by accuracy, with equal accuracy outcomes decided by training time.
Each combination has been tested with 10-fold cross-validation on \dsone (90\% for training and 10\% for testing).

Together with the three types of pre-processing (Boolean conversion, normalization, and standardization) the hyper parameters investigated for each ML algorithm are listed below.
\begin{itemize}
\item \knn: Number of neighbors $[1;30]$.
\item \linreg and \linsvm: Loss function (hinge loss or squared hinge loss).
\item \dt: Criterion to measure quality of split (entropy or Gini impurity), minimal number of samples required for a leaf $[2;12]$, and the maximal depth of the tree $[1;12]$.
\item \dl: Maximal depth of the tree $[1;10]$ (limited for important training time purpose).
\item \rf and \gbdt: Number of estimators $[2;150]$, best parameters of \dt are also used for individual tree.
\item \mlp: Architecture of the network (i.e.~$[1;3]$ hidden layers and ${25,50,100}$ neurons), activation functions (identity, hyperbolic tangent, logistic function, rectified) and solver (adam, sgd or lbfgs).
\item \svm: Kernel used (linear, polynomial, rbf or sigmoid).
\end{itemize}
Note that \nbc and \bnbc do not have hyper-parameters of interest and are only tested for appropriate data pre-processing.

\begin{figure}[t]
\centering
  \includegraphics[width=1.0\linewidth]{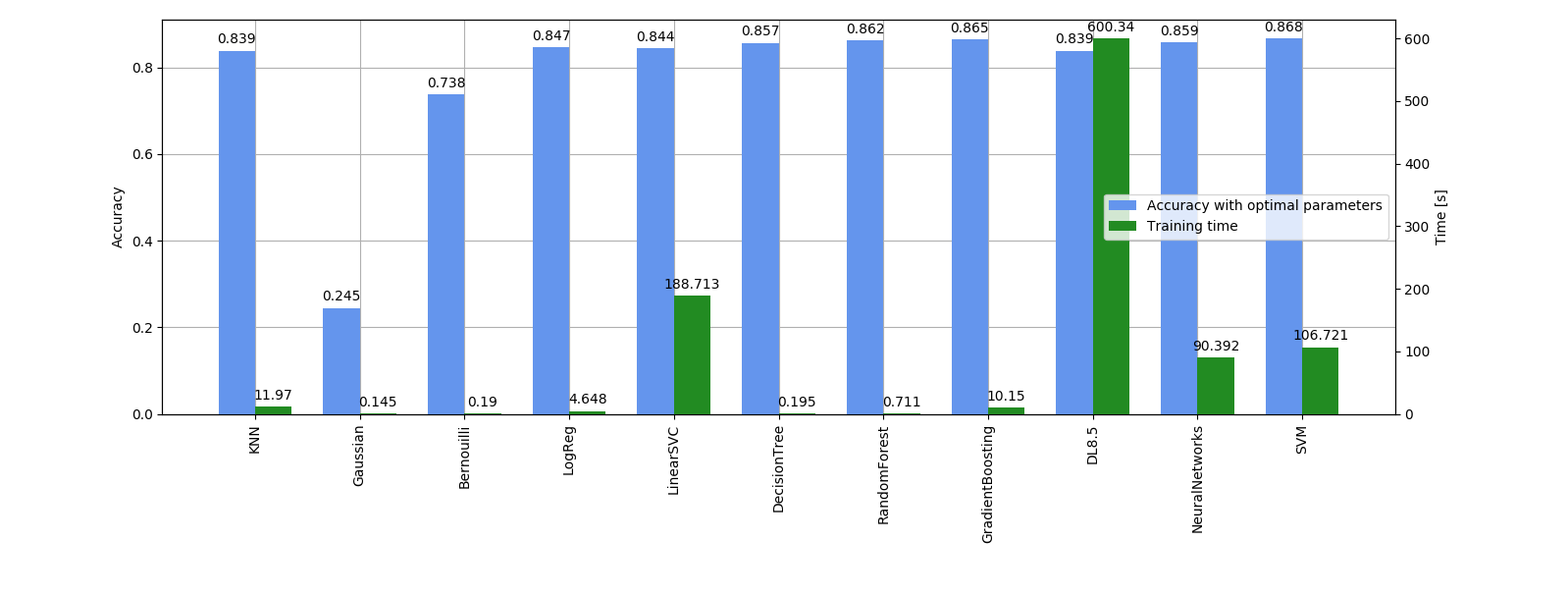}
  \caption{Accuracy and training time for all 11 ML trained classifiers with optimal pre-processing and hyper parameters. 10-fold cross validation has been used with $90\%$-training and $10\%$-testing using the \dsone data set.}
  \label{fig:classifiers_post_algotuning}
\end{figure}

Figure~\ref{fig:classifiers_post_algotuning} shows a summary of different trained classifiers with their best accuracy (according to all the possible pre-processing and hyper-paramater tuning) on \dsone.
This provides a baseline for later comparison as well as an overview of the potential accuracy achievable for each of the 11 ML trained classifiers.
The best choices of data pre-processing and hyper-parameter selection for each ML algorithm as listed below.
\begin{itemize}
\item \knn: Boolean conversion, 16 neighbors.
\item \bnbc: Boolean conversion.
\item \gnbc: Standardization.
\item \linreg: Standardization and square hinge for loss function
\item \linsvm: Boolean conversion and square hinge for loss function
\item \dt: No pre-processing, entropy as criterion to measure quality of split, 10 as minimal number of samples required for a leaf and 6 as the maximal depth of the tree.
\item \dl: Boolean conversion, maximal depth of 10.
\item \rf and \gbdt: No pre-processing, 20 trees as number of estimators.
\item \mlp: Boolean conversion, 50 neurons in the first layer and 100 in the second layer as architecture. Stochastic descent and logistic function as solver and activation function.
\item \svm: Standardization, Gaussian radial basis functions (rbf) as kernel.
\end{itemize}
The above pre-processing and hyper-parameter configurations are used for all later experiments in this paper.

\section{Features Analysis}
\label{sec:FeaturesAnalysis}

This section addresses \textbf{RQ1} (Which features are most significant for packing detection?) by considering the relevance and significance of the 119 different features explored in the literature.
In Biondi et al.~\cite{biondi2019effective} these features are grouped into six categories: Metadata, Section, Entropy, Entry bytes, import functions and Resource features.
See~\ref{sec:appendix_extracted_features} for the full list of features and their categories).
The relevance and significance of the features is considered using three different approaches:
analyzing the weights and choices from the classifiers produced with the ML algorithms (\linreg, \linsvm, \dt, \gbdt, \& \rf);
using iterative K-best feature selection; and
principal component analysis.
The accuracy of the classifier is used to measure the relevance and significance of features, and all analysis in this section is performed on \dsone using 10-fold cross-validation with $90\%$ of \dsone used for training and $10\%$ for testing.

\subsection{Overview of Feature Relevance}
\label{ssec:optimal_features:overview_relevance}

\begin{figure}[p!]
\begin{minipage}[b]{0.48\linewidth}
  \includegraphics[width=1.0\textwidth]{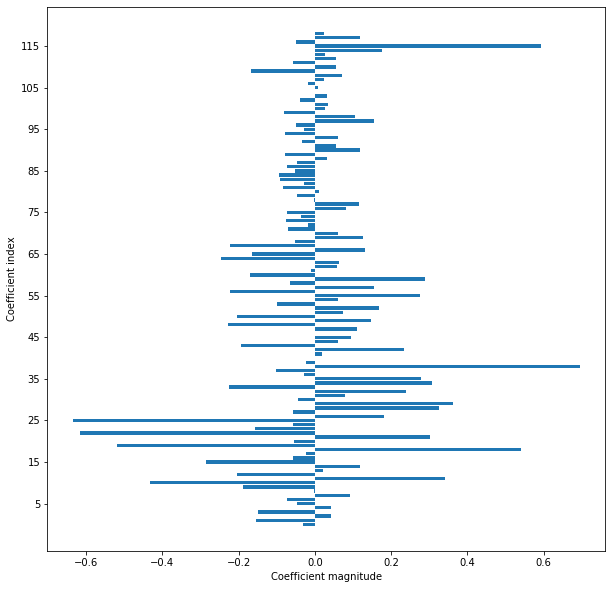}
  \caption{Coefficient magnitudes for the \linreg classifier. 10 most important features by order of relevance are: 39, 116, 19, 30, 12, 29, 35, 22, 60, \& 36.}
  \label{fig:fs_logreg}
\end{minipage}\quad
\begin{minipage}[b]{0.48\linewidth}
 \includegraphics[width=1.0\linewidth]{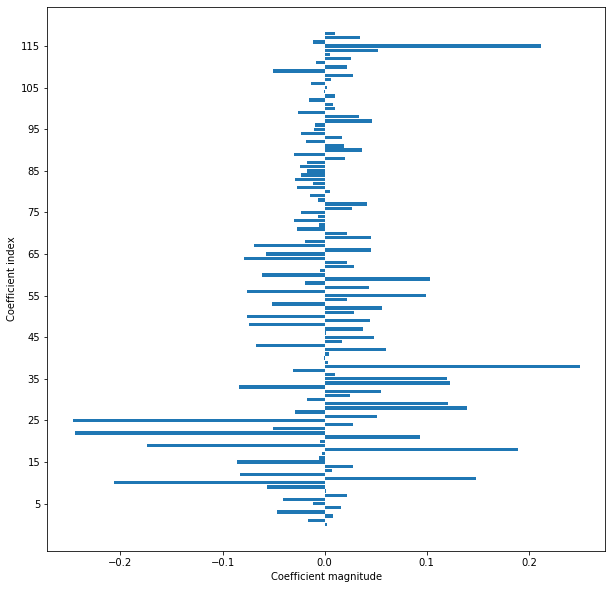}
 \caption{Coefficient magnitudes for the \linsvm classifier. 10 most important features by order of relevance are: 39, 116, 19, 12, 29, 35, 30, 36, 60, \& 56.}
 \label{fig:fs_linearsvc}
\end{minipage}\\
\begin{minipage}[b]{0.48\linewidth}
 \includegraphics[width=1.0\linewidth]{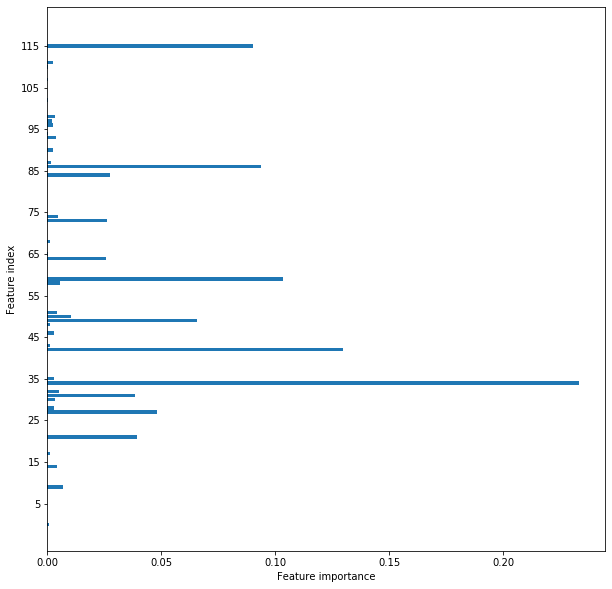}
 \caption{Feature importance for the \dt classifier. 10 most important features by order of relevance are: 35, 43, 60, 87, 116, 50, 28, 22, 32, 85, \& 74.}
 \label{fig:fs_tree}
\end{minipage}\quad
\begin{minipage}[b]{0.48\linewidth}
 \includegraphics[width=1.0\linewidth]{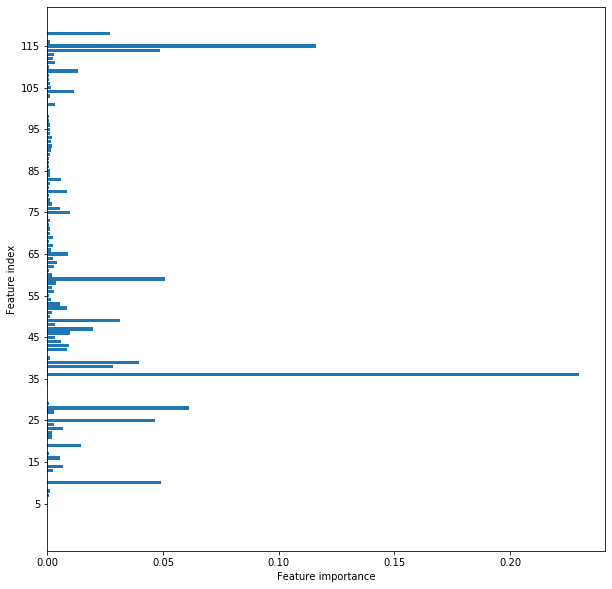}
 \caption{Feature importance for the \gbdt classifier. 10 most important features by order of relevance are: 37, 116, 29, 60, 11, 115, 26, 40, 50, 39, \& 119.}
 \label{fig:fs_gradientboosted}
\end{minipage}
\end{figure}

\begin{figure}[t]
 \centering
 \includegraphics[width=0.48\linewidth]{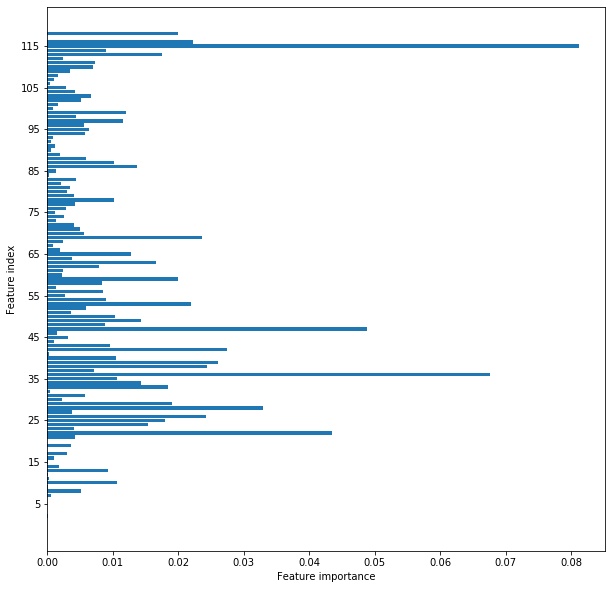}
 \caption{Feature importance for the \rf classifier. 10 most important features by order of relevance are: 116, 37, 48, 23, 29, 43, 40, 39, 27, 70, \& 117.}
 \label{fig:fs_randomforest}
\end{figure}

One approach to understanding which features are most relevant and significant is to examine the classifiers produced by different ML algorithms.
For \linreg, and \linsvm trained classifiers the coefficients magnitude is one way to examine which features are most significant.
For \dt, \rf, and \gbdt the position in the decision tree(s) also implies relevance and significance of features and provides another insight.
Observe that since these two different groups of ML algorithms (\lin-based \lin, \linreg, \&  \linsvm, and \dt-based \dt, \rf, \& \gbdt) operate in different ways, they provide different kinds of insights.

The following features appear relevant and significant to ML classifiers based on this analysis approach (presented in feature index order).
Note that an overview of feature significance for each of the classifiers considered here is presented in~Figures~\ref{fig:fs_logreg} to \ref{fig:fs_randomforest}.
\begin{itemize}
\item Feature 19 corresponds to the \emph{size of the stack to reserve}. Since unpacking implies in-memory operations and execution, this feature appears to be intrinsically linked to packing behavior. This feature is mainly significant in \lin-based models.
\item Feature 29 corresponds to the \emph{number of readable and writable sections the PE holds}. Typically, a packer will unpack its code in specific sections which means it will re-write some of its previously encrypted sections. Therefore it is expected that this feature is in the top ten used features of \linreg, \linsvm, \rf and \gbdt.
\item Feature 35 corresponds to the \emph{entry point not being in a standard section}. This can could happen when the entry point which corresponds to the unpacking stub is in a dedicated section. This feature is in the top ten features of \linreg, \linsvm and \dt.
\item Feature 37 corresponds to the \emph{ratio between raw data and virtual size for the section of the entry point}. A huge ratio can be link to a section that will be modified during execution, in the case of the entry point section this is quite suspicious. This feature is one of the top features of \rf and \gbdt.
\item Feature 39 corresponds to the \emph{number of sections having their virtual size greater than their raw data size}. A packed executable will typically create this change in section size due to modifying data inside a section. This is a top ten features of \linreg, \linsvm, \rf and \gbdt.
\item Feature 43 corresponds to the \emph{byte entropy of code (.text) sections}. Since packing implies generally compression/encryption, entropies of code sections is generally high. This explains why this feature is often used in literature~\cite{DBLP:journals/ieeesp/LydaH07,choi2008pe,jeong2010generic,raphel2015information,ugarte2014adoption}. 
\item Feature 60 represent the \emph{12th-byte value following the entry point}. The significance of this byte specifically is interesting to observe and is in the decision process for \linreg, \linsvm and \dt.
\item Features 115-116 are related to \emph{the number of API calls imported} by the binary and \emph{often used in malicious software} according to~\cite{zakeri2015static}. Feature 115 corresponds directly to the number of API calls imported while Feature 116 is related to the ratio between malicious API calls imported and the total API calls imported. These API calls are not intrinsically malicious but allow easily alteration or obfuscation of control flow of a running program. This can be by allocating memory to store unpacked files, getting address of specific API function call after unpacking, etc. Feature 116 is part of top ten used features of all explored methods while Feature 115 is also utilized in \gbdt.
\end{itemize}

\subsection{Selecting Features}
\label{ssec:optimal_features:selection}

Another approach to better understand the relevance and significance of features is to explore how to reduce the features used by some of the ML algorithms.
Here this is done in two ways:
the first is by exploring the \emph{$K$-best} features extracted from the initial models; and
the second is by \emph{iterative} increase where (increasing iteratively) $K$ features are chosen and the intersection of the best features is kept over iterations.

To select the $K$-best features for models, the \emph{scikit} implementation (namely \emph{SelectFromModel} function) was used. A threshold is used to decide if a feature is sufficiently important to be kept, here the threshold has been incremented progressively, and all feature combinations tested with combinations that did not decrease accuracy of the classifier by more than 5\% kept.
For the list of combinations of features satisfying this requirement, an indicative \emph{time-to-accuracy} ratio has been computed which represents the ratio between the percentage of time saved and the percentage of accuracy lost. Observe that this process of selection is only possible on estimators making direct use of coefficient or feature importance namely: \linreg, \linsvm, \dt, \rf and \gbdt.

\begin{table}[t]
\centering
\resizebox{\textwidth}{!}{%
\begin{tabular}{|c|c|c|c|c|c|c|c|}
 \hline
 Classifier & Selection & Best \# features & Old accuracy & New Accuracy & Old time & New time & Ratio\\
 \hline\hline
 \linreg & Iterative & 103 & 0.8476 & 0.8480 & 4.385 & 3.462 &(-)446  \\
 \hline
 \linsvm & Iterative & 84 & 0.8440 & 0.8470 & 281.243  & 128.804 &(-)152 \\ 
 \hline
 \dt & K best & 16 & 0.8572 & 0.8556 & 0.1797 & 0.0218 & 470 \\ 
 \hline
 \rf & Iterative & 24 &0.8627 & 0.8622 &0.6423 &0.2843 & 961.68  \\ 
 \hline
 \gbdt & K best & 28 &0.8657 & 0.8656 & 9.6522 & 2.6470 & 6283  \\ 
 \hline
\end{tabular}%
}
\caption{Best results for feature selection based on ratio value.}
\label{Tab:fs_table}
\end{table}

A summary of the feature relevance process outcomes can be seen in Table \ref{Tab:fs_table}.
These results indicate that feature selection is in general beneficial for every tested ML classifier measured by the high value of time-to-accuracy ratios
For ML approaches such as \linsvm  which take a significant training time, feature selection not only reduces training time but also improves classifier accuracy.
A small improvement in performance is also observed for \linreg.
Generally the results indicate that \lin-based models tend to use more features than decision tree based models. 
This confirms observations in Section \ref{ssec:optimal_features:overview_relevance}.

\subsection{Principal Component Analysis}
\label{ssec:PCA}

\begin{table}[t]
\centering
\resizebox{\textwidth}{!}{%
\begin{tabular}{|c|c|c|c|c|c|c|}
\hline
Classifier & Best \# components & Old accuracy & New accuracy & Old time & New time (s) & Ratio\\
\hline\hline
\knn & 71 & 0.8391 & 0.8411  & 11.97 & \textbf{0.634462} & (-)397  \\
\hline
\gnbc & 14 & 0.2458 & \textbf{0.8086 } & 0.145 & 0.0201 & (-)0.37 \\ %
\hline
\linreg & 99 & 0.8476 & 0.8513  & 4.648  &3.808 & (-)41\\
\hline
\linsvm & 101 & 0.844 & 0.8458  & 188.713 & 153.405 & (-)87.72\\
\hline
\dt & 58 & 0.8572 & 0.8464  & 0.195 & 0.0829 & 46\\
\hline
\rf & 11 & 0.8627 & 0.8589 & 0.711 & 0.6717 & 13 \\
\hline
\gbdt & 18 & 0.8657 & 0.8622  & 10.15 & 5.13846  & 122.33 \\
\hline
\mlp & 57 & 0.8592 & 0.8628  & 90.392 & \textbf{85.882} & (-)12 \\
\hline
\svm & 101 & 0.8682 & 0.8682  & 106.72  & \textbf{92.990} & 1117\\
\hline
\end{tabular}%
}
\captionsetup{justification=centering}
\caption{Top result from applying PCA and keeping only iterations improving time and finally sorting them by accuracy.}
\label{Tab:pca_table}
\end{table}

\emph{Principal component analysis} (PCA) aims to reduce the dimension of a feature space by grouping features together to constitute new components. Each new component is an independent novel linear combination of some prior features. These are ranked by importance, the first component trying to capture as much information (i.e.~variance of the data) as possible.

Using different number of principal components, best sets of these new features in terms of accuracy were selected.
Table~\ref{Tab:pca_table} summarizes best results from applying PCA over \dsone before training of our ML classifiers. Note that since PCA expects standardization as a part of the process, \dl and \bnbc are ineligible since they require binary data.

An overview of the results of PCA can be seen in Table~\ref{Tab:pca_table}.
Although computation times generally improve in every case, three types of results could be distinguished.
First, for \knn and \gnbc the number of features is substantially decreased and a significant improvement can be observed in accuracy.
These methods being more based on distance and statistical distribution, it appears intuitive creating principal components will benefit to them.\\
Second, the impact of PCA on algorithms like \linreg, \linsvm, \mlp and \svm seems less significant: the number of features didn't decrease significantly, and accuracy performance did not reveal substantial improvement.
These algorithms already attach importance to different feature and since PCA reduces dimensionality by employing linear combinations over data, these are somewhat redundant in their effect. 
Although this limited the improvements in accuracy, improvement in training time can be observed for \mlp and \svm.
Third, tree-based ML algorithms such as \dt, \rf and \gbdt dramatically decreases the number of features without impacting accuracy.
Since trees-based algorithms derive their cogency from the diversity of features to perform their split, applying PCA reduces dimensionality and creates more complex feature to split upon for tree-based models.

Ultimately, PCA will be kept as a step for later application of \knn, \gnbc, \mlp and \svm since improvements in accuracy or training time justifies always using PCA.

\subsection{Accuracy with Modified Features}
\label{ssec:RQ1_accuracy}

\begin{figure}[t]
\centering
  \includegraphics[width=1.0\linewidth]{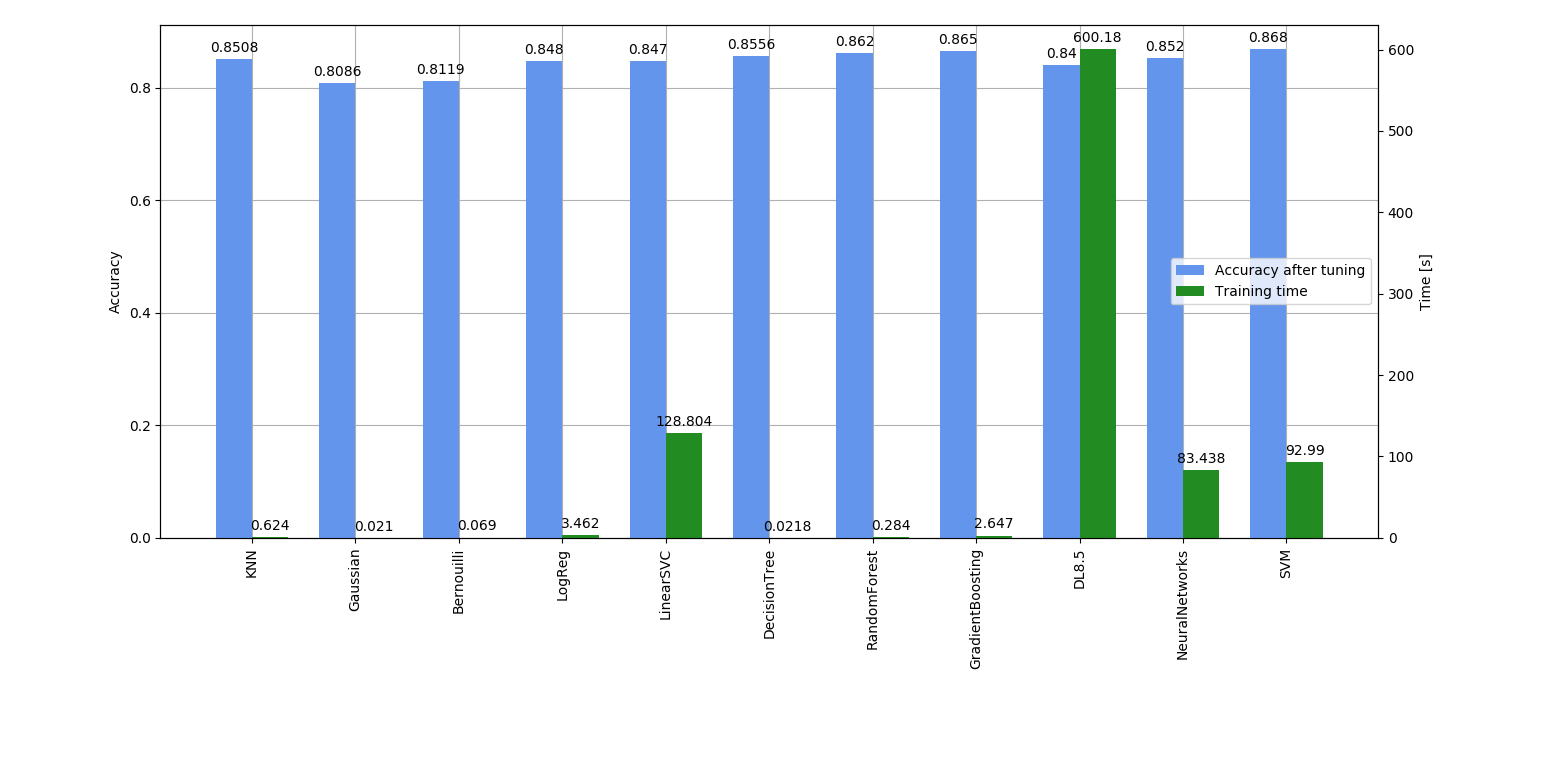}
  \caption{Accuracy and training time for all ML classifiers after processing of features.}
  \label{fig:classifiers_post_datatuning}
\end{figure}

Figure~\ref{fig:classifiers_post_datatuning} illustrates the accuracy and training time of all ML classifiers after applying the best choices from the above analysis. 
Feature selection is applied to \linreg, \linsvm, \dt, \rf and \gbdt, and PCA is applied to \knn, \gnbc, \mlp and \svm.
Using feature selection/PCA reduces the number of necessary features stored and used for training. Most ML algorithms can significantly reduce their number of features while maintaining their accuracy.

\medskip

Regarding \textbf{RQ1}, we observe that a small number of features are clearly more influential than others. These are: the stack reserve (Feature 19), number of readable and writable sections (Feature 29), non-standard entry point section (Feature 35), raw and virtual data size of sections (Features 37 \& 39), byte entropy of \texttt{.text} section (Feature 43), 12th entry point byte (Feature 60), and malicious API calls (Features 115-116). In \lin classifiers, all features have weighted contributions to the classification process even if some of them are significantly lower weight than others. By contrast, in tree-based models, classifiers take advantage of a small amount of features to classify a sample. These properties of the classifiers and the diversity of the features and their provenance correlates with prior results showing different features and approaches
can be effective and play a role in classification even if many are not significant independently.

\section{Classifier Effectiveness}
\label{sec:Effectiveness}

This section addresses \textbf{RQ2} (Which ML classifiers are effective for packing detection?) by examining various metrics for classification over a large data set.
The adaptability of the classifiers is also evaluated by testing the trained classifiers on specific packers.

\subsection{Evaluation of Classification Metrics}
\label{ssec:metrics}

\begin{figure}[t]
\centering
  \includegraphics[width=0.90\linewidth]{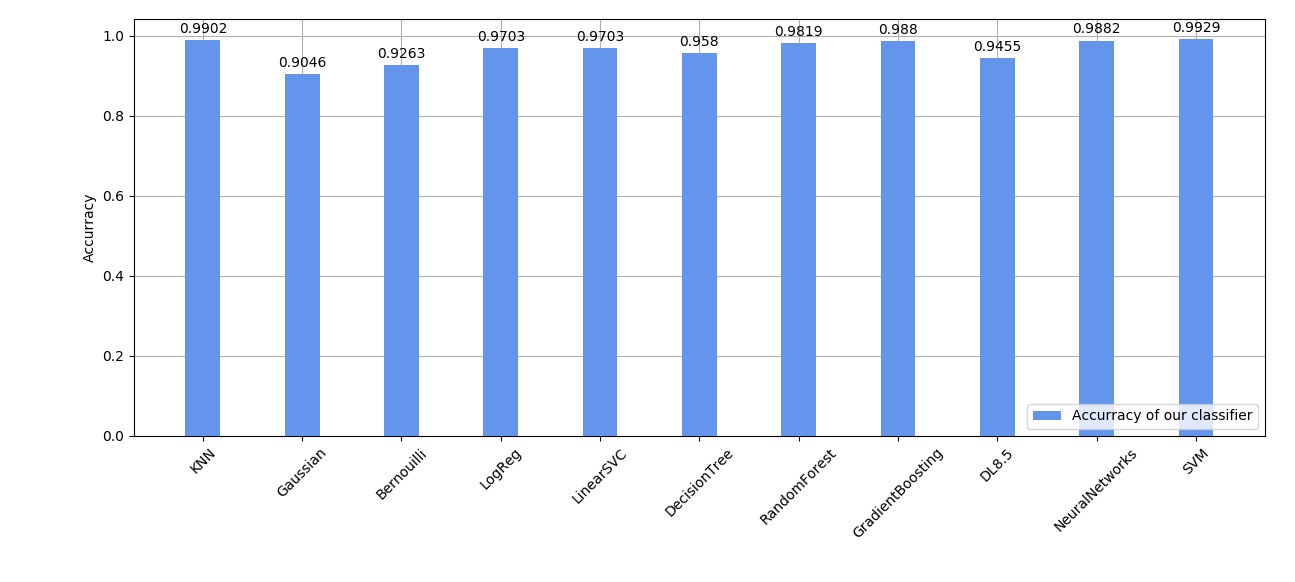}
  \caption{Accuracy for all 11 ML trained classifiers after tuning tested on \dstwo.}
  \label{fig:assertion_test}
\end{figure}

To assess the performance of the 11 ML trained classifiers an analysis is performed over the larger \dstwo.
For the results in this section $10$-fold cross-validation is performed with $90\%$ of \dstwo used for training and $10\%$ for validation.

An overview of the average accuracy for each classifier can be seen in Figure~\ref{fig:assertion_test}.
Observe that in all cases except for \bnbc the ML classifier accuracy is higher than when performing the parameter tuning (see Section~\ref{ssec:RQ1_accuracy} and Figure~\ref{fig:classifiers_post_datatuning} for details).
These results indicate that the choice of parameters appears reasonable and that they are not over-fitted to the data set used for the parameter tuning (\dsone).

\begin{table}[t]
     \centering
     \resizebox{\textwidth}{!}{%
     \begin{tabular}{|c||c|c|c|c|c|c||c|c|c|}
     \hline
     Classifier & precision$_{p}$ & precision$_{np}$ & recall$_{p}$  & recall$_{np}$  & $F$-score$_{p}$ & $F$-score$_{np}$ & precision$_{wa}$ &  recall$_{wa}$ & $F$-score$_{wa}$\\
     \hline\hline
     \knn & 0.9746 & \textbf{0.9955} & \textbf{0.9867} & 0.9913 & 0.9806 & 0.9934 & 0.9903 &  0.9902 &   0.9902\\
     \hline
     \gnbc & 0.7882 & 0.9478 & 0.8487 & 0.9234 & 0.8173 & 0.9354 & 0.9077 & 0.9046 & 0.9057\\
     \hline
     \bnbc & 0.8091 & 0.9728 & 0.9219 & 0.9278 & 0.8619 & 0.9498 & 0.9321  &0.9264 &0.9279 \\ 
     \hline
     \linreg & 0.9415 & 0.9801 & 0.9407 & 0.9803 & 0.9411 & 0.9802 & 0.9704  & 0.9703  & 0.9703\\ 
     \hline
     \linsvm & 0.9393 & 0.9809 & 0.9432 & 0.9795 & 0.9412 & 0.9802 & 0.9704 & 0.9704 & 0.9704\\ 
     \hline
     \dt & 0.9119 & 0.9738 & 0.9225 & 0.9700 & 0.9172 & 0.9719 & 0.9586 &  0.9581 &   0.9582\\
     \hline
     \rf & 0.9733 & 0.9847 & 0.9543 & 0.9912 & 0.9637 & 0.9880 & 0.9819 &0.9820   & 0.9819\\
     \hline
     \gbdt & 0.9844 & 0.9892 & 0.9676 & 0.9948 & 0.9759 & 0.9920 & 0.9880 & 0.9880 &0.9880 \\
     \hline
     \dl & 0.9473 & 0.9451 & 0.8276 & 0.9847 & 0.8834 & 0.9645 & 0.9456 & 0.9456 & 0.9443\\
     \hline
     \mlp & 0.9729 & 0.9934 & 0.9805 & 0.9908 & 0.9767 & 0.9921& 0.9882 & 0.9882 & 0.9882\\
     \hline
     \svm & \textbf{0.9875} & 0.9947 & 0.9843 & \textbf{0.9958} & \textbf{0.9859} & \textbf{0.9952} & \textbf{0.9929} & \textbf{0.9929} & \textbf{0.9929}\\
     \hline
     \end{tabular}
    }
     \caption{Detailed metrics for all 11 ML trained classifiers after tuning tested on \dstwo.}
    \label{Tab:classifier_metrics}
\end{table}

A more detailed exploration of the metrics for evaluation the different classifiers is presented in Table~\ref{Tab:classifier_metrics}.
For each classifier the average of the following metrics is presented.
\begin{itemize}
\item The \emph{precision} of a class $A$ is defined as the ratio of samples correctly labeled $A$ over all samples labeled $A$. 
\item The \emph{recall} of a class $A$ is defined as the ratio of samples correctly classified as $A$ over all samples belonging to class $A$.
\item The \emph{$F$-score} of a class $A$ is defined as $\frac{2 \cdot \textrm{precision}_{A} \cdot \textrm{recall}_{A}}{\textrm{precision}_{A} + \textrm{recall}_{A}}$.
\end{itemize}
The subscript $p$ denotes the packed class, the subscript $np$ denotes non-packed class, and the subscript $wa$ denotes the weighted average.

Overall the precision, recall, and $F$-score tend to be higher for the non-packed class for almost all ML classifiers.
The only exception is small improvement for \dl regarding precision.
Globally of the 11 ML trained classifiers \svm scores the best in all metrics except for precision$_{np}$ and recall$_p$ which are both highest with \knn.

Overall the Bayes-based classifiers (\gnbc \& \bnbc) perform relatively poorly compared to the other ML trained classifiers.
The significant weakness of these classifiers appears to be in their ability to detect packed samples, with all metrics much lower than non-packed metrics.
\gnbc performs worse than \bnbc, which is in line with the prior tuning results.

The linear models (\linreg \& \linsvm) performed in the middle compared with the other algorithms, and very similarly to each other.
There is no easy way to distinguish \linreg from \linsvm fro these metrics, and so a decision between them may consider the training time where \linreg is much faster.

The decision tree based classifiers (\dt, \gbdt, \rf, \& \dl) mostly perform well, with \gbdt performing slightly better than \rf while \dt and \dl are less effective on all metrics.

\mlp appears effective and comparable with the best decision tree based classifier (\gbdt) although worse than \knn and \linsvm.

Finally, \svm appears to offer the best overall performance among the different classifiers.

\subsection{Adaptability}
\label{ssec:specific_packers}

\begin{table}[t]
     \centering
     \resizebox{\textwidth}{!}{%
     \begin{tabular}{|c|c|c|c|c|c|c|c|c|c|c|}
     \hline
     \multirow{2}{*}{Classifier} & \multicolumn{2}{c|}{UPX} & \multicolumn{2}{c|}{kkrunchy} &\multicolumn{2}{c|}{MPress} & \multicolumn{2}{c|}{TElock} & \multicolumn{2}{c|}{PEtite}\\
     \cline{2-11}
     & \dstwo & \extendset$_{UPX}$ &\dstwo & \extendset$_{kkrunchy}$ &\dstwo & \extendset$_{MPress}$ &\dstwo & \extendset$_{TElock}$ &\dstwo & \extendset$_{PEtite}$ \\
     \hline\hline
     \knn & 0.9873 & 0.9756 & 0.7682 & 0.968 & 0.9264 & 0.9728  & 0.7606 & 0.8359 & 0.016 & 0.9277\\
     \hline
     \gnbc & 0.9873 & 0.9955 &  0.5133 & 1.0 & 0.3676 & 0.9689 & 0.3204 & 0.3735 & 0.0115 & 0.0246\\
     \hline
     \bnbc & 0.9924 & 0.9822 & 0.7433 & 0.972 & 0.875 & 0.9689 & 0.5922 & 0.9486 & 0.1598 & 0.7701\\
     \hline
     \linreg & 0.9600 & 0.9977  & 0.792 & 0.988 & 0.5755 & 0.9844 & 0.1837 & 0.4940 & 0.0098 & 0.8998\\
     \hline
     \linsvm & 0.9623 & 0.9977 & 0.816 & 0.982 & 0.6337 & 0.982 & 0.1739 & 0.8675 & 0.009 & 0.8456\\
     \hline
     \dt & 0.9911 & 0.9955  & 0.01 & 0.12 & 0.9709 & 0.9903 & 0.426 & 0.4466 & 0.003 & 0.08\\
     \hline
     \rf & 0.9955 & 1.0  & 0.34 & 1.0 & 0.9709 & 0.9903 & 0.1482 & 0.7035 & 0.0 & 0.0476\\
     \hline
     \gbdt & 0.9955 & 1.0 & 0.002 & 0.998 & 0.9709 & 1.0 & 0.3913 & 0.91 & 0.0 & 1.0\\
     \hline
     \dl & 0.75 & 0.8824 & 0.03 & 0.836 & 0.93 & 0.9573 & 0.6267 & 0.1660 & 0.0268 & 0.267\\
     \hline
     \mlp & 0.9848 & 1.0  & 0.5989 & 1.0 & 0.9166 & 0.9883 & 0.5415 & 0.7509 & 0.0089 & 0.9688\\
     \hline
     \svm & 0.9671 & 1.0 & 0.037 & 0.996 & 0.897 & 1.0 & 0.6227 & 0.8992 & 0.0089 & 0.9852\\
     \hline
     \end{tabular}
    }
     \caption{Accuracy for all 11 ML trained classifiers after tuning applied to a \dstwo training set and \extendset training set with added samples.}
    \label{Tab:perf_various_packers}
\end{table}

In the prior experiments the data sets consist of collected samples labeled with majority vote (as detailed in Section~\ref{ssec:sets}).
To explore issues related to poorly represented packers and adaptability of the ML algorithms, the accuracy contrasting \dstwo with \extendset$_{f}$ is here evaluated with a variety of packers $f$.
This contrast is detailed for five packer families: UPX, kkrunchy, MPress, TElock and PEtite.
In each case all 11 ML trained classifiers are trained on both \dstwo and \extendset$_{f}$ for each $f$ of the five packers and evaluated on \dstwotest$_{f}$.

These five packers are chosen as they represent very different frequencies within \dstwo.
The frequency of UPX is 11978 in \dstwo and 12478 in \extendset$_{UPX}$.
The frequency of kkrunchy is 71 in \dstwo and 571 in \extendset$_{kkrunchy}$.
The frequency of MPress is 344 in \dstwo and 844 in \extendset$_{MPress}$.
The frequency of TElock is 351 in \dstwo and 851 in \extendset$_{TElock}$.
Finally, the frequency of PEtite is 23 in \dstwo and 523 in \extendset$_{PEtite}$.
Note that the frequencies for all packers except $f$ in \extendset$_{f}$ remains unchanged.

The experimental approach is an adaptation of the approach used in~\cite{aghakhani2020malware}.
All the ML classifiers are trained using either \dstwo or \extendset$_{f}$ and then evaluated on \dstwotest$_{f}$.
The goal of this experiment was to observe the adaptability of the ML trained classifiers to detect new samples created with known packer $f$ in \extendset$_{f}$ and compare this with the ML classifier that is trained on \dstwo without the additional packed samples of family $f$.
As usual all experiments were performed with $10$-fold cross-validation, although here the training set and testing sets are distinct. %

An overview of the results can be seen in Table~\ref{Tab:perf_various_packers}.
Observe that in almost all cases the extended training set improves the accuracy for classification on \dstwotest$_{f}$ for all $f$ (the only exception is TElock using \mlp).
Observe that not all classifiers are equal against new packed samples and many perform poorly on unknown packers.
Since \knn keeps and uses all the data set for its model, more new samples appear necessary to improve classification to a high accuracy.
\gnbc seems to perform great on samples sufficiently present in the data set and show difficulties facing other packers (even when extending the data set).
Comparatively, \bnbc performs better by directly recognizing more packers and being able to integrate new information for its model.
\linreg and \linsvm models both perform well and are able to integrate new samples to improve their model (particularly \linsvm).
\dt and \dl performs well for some packers (UPX and MPress) without additional samples but fail to integrate new packers efficiently even when extending the training set.
Contrarily, \gbdt adapts extremely well and \rf also improves, albeit less than \gbdt.
\dl shows improvement with new information, although like \rf  struggling to handle PEtite.
\mlp shows excellent results by detecting many packers with high accuracy and integrating new information effectively.
Finally, \svm already performed extremely like \mlp and integrated new information even more effectively. %

\medskip

These results address \textbf{RQ2} by examining the effectiveness of all 11 ML algorithms via various metrics and experiments.
There are 4 ML algorithms that appear to be most effective as briefly highlighted below.
\begin{itemize}
\item \knn is memory expensive (all data points are part of the model) and requires many samples to improve its model. However, \knn avoids false positives/negatives and performs well if samples are sufficiently present in the training data set.
\item  \gbdt is light in training (compared to other methods). \gbdt offers good performance and can easily extend its model in case of a new packers detected (if samples are provided). (Note \rf is slightly less efficient but in many ways comparable to \gbdt if a decision trees based classifier is desirable.)
\item  \mlp is resource demanding for training but offers good performance and adaptability.
\item \svm appears to be the most effective and adaptable ML algorithm for packing detection. \svm has a low false positive/negative rate and all packers are detected in the adaptability experiments.
\end{itemize}


\section{Economical Analysis}
\label{sec:Economics}

\begin{table}[t]
     \centering
     \begin{tabular}{|c|c|c|c|c|c|}
     \hline
     Classifier & Baseline & Period 1 & Period 2  & Period 3  & Period 4\\
     \hline\hline
     \knn & 0.9902 & 0.9899 & 0.9830 & 0.9823 & 0.9696 \\
     \hline
     \gnbc & 0.9046 & 0.9034 & 0.8940 & 0.8738 & 0.8751 \\
     \hline
     \bnbc & 0.9263 & 0.9170 & 0.9158 & 0.8890 & 0.8545 \\
     \hline
     \linreg & 0.9703 & 0.9634 & 0.9506 & 0.9452 & 0.9402 \\
     \hline
     \linsvm & 0.9703 & 0.9627 & 0.9509 & 0.9466 & 0.9396 \\
     \hline
     \dt & 0.9580 & 0.9551 & 0.9438 & 0.9395 & 0.9324  \\
     \hline
     \rf & 0.9819 & 0.9776 & 0.9751 & 0.9768 & 0.9685 \\
     \hline
     \gbdt & 0.9880 & 0.9851 & 0.9806 & 0.9824 & 0.9701 \\
     \hline
     \dl & 0.9456 & 0.9383 & 0.9408 & 0.9037 & 0.9016 \\
     \hline
     \mlp & 0.9882 & 0.9823 & 0.9781 & 0.9757 & 0.9669 \\
     \hline
     \svm & 0.9929 & 0.9924 & 0.9922 & 0.9913 & 0.9788 \\
     \hline
     \end{tabular}
     \caption{Economical analysis for all 11 ML trained classifiers after tuning trained on \dstwo and tested on \dstwo (baseline) and \dsthree split by chronological period.} 
    \label{Tab:time_analysis}
\end{table}

\begin{table}[t]
     \centering
     \resizebox{\textwidth}{!}{%
     \begin{tabular}{|c|r|r|r|r|r|r|r|}
     \hline
     Classifier & Train time [s] & Uptime $0.92$ [s] & Ratio $0.92$ & Uptime $0.95$ [s] & Ratio $0.95$ & Uptime $0.97$ [s] & Ratio $0.97$\\
     \hline\hline
     \knn & 2.1259 & 8,986,447 & 4,227,126 & 6,846,280 & 3,220,415 & 4,907,050 & 2,308,222\\
     \hline
     \gnbc &0.0368 & N/A & N/A  & N/A & N/A & N/A & N/A\\
     \hline
     \bnbc & 0.1673 & 1,661,053 & 9,928,592  & N/A & N/A & N/A & N/A\\
     \hline
     \linreg &11.0572 &8,346,239 & 754,824  & 2,791,949 & 252,500 & 134,757 & 12,187\\
     \hline
     \linsvm & 316.172 & 9,961,820 &31,507  & 2,865,875 & 9,064 & 90,561 & 286\\
     \hline
     \dt & 0.0607 & 6,874,338 & 113,251,038  & 1,680,899 & 27,691,918 & N/A & N/A \\
     \hline
     \rf &0.7492 & 14,776,270 & 19,722,731   & 9,715,226 & 12,967,467 & 4,812,299 & 6,423,250\\
     \hline
     \gbdt & 2.7977 & 10,303,293 &3,682,772  & 7,616,469 & 2,722,403 & 5,116,627 &1,828,869\\
     \hline
     \dl & 599.31 & 3,310,904  & 5,524  & N/A & N/A & N/A & N/A\\
     \hline
     \mlp & 317.29 & 12,244,014 & 38,589 & 8,004,772 & 25,228 & 4,397,608 & 13,859\\
     \hline
     \svm & 161.9987 & 9,262,621 &57,177 & 7,473,608  & 46,133 & 5,900,674 & 36,424\\
     \hline
     \end{tabular}
    }
     \caption{Economical analysis for all 11 ML trained classifiers after tuning trained on \dstwo including time to train, uptime while maintaining $F$-scores, and uptime to train time ratio for $F$-scores $0.92$, $0.95$, and $0.97$.}
    \label{Tab:time_analysis_fscore}
\end{table}

This section addresses \textbf{RQ3} (Which ML algorithms perform well in long term for packing detection?) by exploring the effectiveness of all 11 ML algorithms over chronological data.

This section evaluates the effectiveness of all 11 ML algorithms over time and the retraining required to maintain a high level of accuracy.
To perform this analysis all 11 ML algorithms are trained on \dstwo and their effectiveness evaluated on \dsthree.
\dstwo is used to train each classifier, thus training on samples collected from 2019-10-01 to 2020-02-28.
The testing is then done on \dsthree which corresponds to malware samples obtained from 2020-04-01 to 2020-05-26.
\dsthree has been divided into four subsets corresponding to separate periods of time as follows.
Period 1 designates malware obtained from 2020-04-01 to 2020-04-14.
Period 2 malware samples are obtained from 2020-04-15 to 2020-04-28.
Period 3 malware samples are obtained from 2020-04-29 to 2020-05-12.
Finally, period 4 malware samples are obtained from 2020-05-13 to 2020-05-26

An overview of the accuracy and efficiency of all 11 ML trained classifiers on these four periods from \dsthree are shown in Table~\ref{Tab:time_analysis}.
As expected, the effectiveness of all 11 ML trained classifiers decreases with time.
This corresponds to prior research \cite{biondi2019effective} and can be explained by different trends in packers used depending on the period of interest.
Some packers are employed frequently by malware in the period 1 of \dsthree and are slowly replaced by other packers during later periods.

The training time and uptime can be used to predict the economics of an ML trained classifier by calculating the retraining required to maintain a given quality.
Here the quality is picked as a target $F$-score and then the uptime is the time after training during which the ML trained classifier has at least the chosen quality.
An overview of the quality measures for three different $F$-scores ($0.92$, $0.95$, and $0.97$) are shown in Table~\ref{Tab:time_analysis_fscore}.
Using the period calculated $F$-scores for each period a curve for more precise estimates is obtained by a quadratic least squares regression. %
This curve can then be used to predict the time that a particular trained classifier will drop below a given $F$-score threshold.
The ratio then indicates the economics of cost in training to uptime, with higher numbers being better,
Here these indicate that \dt and \rf have the best economics due to their extremely quick training time and generally high accuracy and $F$-score.
\knn and \gbdt also perform well, although the significant increase in training time brings their economic ratio down significantly.
\linreg, \linsvm, \mlp, and \svm are all also capable of meeting the requirements, but are economically hampered by their significantly larger training times.
However, observe that since the uptime for \knn, \rf, \gbdt, \mlp, and \svm all have uptimes $> \sim 80$ days even the $\sim 5$ minutes training time for the highest (\mlp) is not too significant.

In machine learning, the training data set used is critical to ensure good performance (as noted also in Section~\ref{ssec:specific_packers}).
Although the performance of all classifiers decreases over time, some classifiers are more robust than others and require less training time and frequency to maintain a given level of effectiveness.
This allows us to answer \textbf{RQ3}:
\knn, \linreg, \dt, \rf, \gbdt, \mlp and \svm are all able to maintain an $F$-score of $0.95$.
From the pure economics, \dt and \rf significantly out perform the others, although considering all training times as reasonable then \knn, \rf, \gbdt, \mlp, and \svm are all able to remain effective for long periods of time between retraining.

\section{Challenges to Validity}
\label{sec:validity}

This section briefly discusses the main potential challenges to the validity of this work and a brief examination of the significance of dynamic features in packer detection.

One challenge to the validity of many works in this area is the selection of ground truth. 
The approach used here has the advantage of a broad selection of samples that are believed to be malware.
Although this has the potential limitation of having few or no clean samples, this choice was made since this work explores the role of packing detection within a malware detection setting.
Another challenge the ground truth used here is that this relies upon a multiple approaches that in turn are subject to their own limitations.
Broadly the consensus approach used here should mitigate some of the limitations (see also \cite{biondi2019effective}). However, this leaves those approaches and this work relatively vulnerable to rare or unknown packers.
To some extent the adaptability exploration (see Section~\ref{ssec:specific_packers}) addresses this, by showing the effectiveness of the ML algorithms when presented with new training data.
The above aside, clearly better ground truth and training data should improve the behaviour of all the ML algorithms considered here.

Another challenge is the use of only accuracy as a metric for the tuning and most analysis performed in this work. 
This choice was to simplify the work by having a single metric (that in turn accounts for both positive and negative outcomes) without giving too much bias or favour to any particular outcome.
The exploration of different metrics (see Section~\ref{ssec:metrics}) indicates that other metrics do not vary significantly from accuracy.
Thus, the results here should be easily broadly applicable to other metrics, and the methodology easily adaptable should another metric be more important in another application.

A further challenge is the use of only static features for packing detection although some related works have also considered dynamic features \cite{bat2013dynamic,DBLP:conf/ccs/Cheng0FPCZM18,kang2007renovo}.
The problem of packing detection as part of a larger malware detection and defence system motivated the choice to focus only on static features in this work, as well as this being proven highly effective both here and in related works.
However, a small investigation was done to explore whether dynamic features may have been significantly useful. This is included below, although due to the cost of feature extraction for dynamic features this work chose to focus on larger data sets and extensive results on cheap features to extract.

\subsection{A Note on Dynamic Features}
\label{sssec:dynamic}

\begin{table}[t]
     \centering
     \begin{tabular}{|c|c|c|}
     \hline
     Classifier & Static Features Only & Static \& Dynamic Features \\
     \hline\hline
     \knn & 0.9889 & 0.9889 \\
     \hline
      \gnbc & 0.939 & 0.943 \\
     \hline
     \bnbc & 0.947 & 0.947 \\
     \hline
     \linreg & 0.9836 & 0.9836\\
     \hline
     \linsvm & 0.9833 & 0.9833\\
     \hline
     \dt & 0.9864 & 0.9864 \\
     \hline
     \rf & 0.9888 & 0.9888 \\
     \hline
     \gbdt & 0.9941 & 0.9941 \\
     \hline
     \dl & 0.9906 & 0.9907 \\
     \hline
     \mlp & 0.9902 & 0.9915\\
     \hline
     \svm & 0.9938 & 0.9931\\
     \hline
     \end{tabular}
     \caption{Accuracy comparison for all 11 ML algorithms using \dsthree static features and with additional dynamic features.}
    \label{Tab:perf_dynamic_feature}
\end{table}

Since dynamic features have been used in the literature, and since packing could be more obvious when a program is executed, this raises a potential further research questions.
\textbf{RQA Do dynamic features improve packing detection?}
This section briefly presents results exploring this question.

Dynamic entropy features are often used in the literature to detect packing \cite{bat2013dynamic,bat2017entropy,jeong2010generic}.
To explore whether dynamic features were significant, we considered four dynamic entropy features.
The \emph{maximum increase of entropy of a section} (Feature D1).
The \emph{maximum decrease of entropy of a section} (Feature D2).
The \emph{biggest change of entropy of the whole file} (Feature D3).
The \emph{change of entropy of the entrypoint section} (Feature D4).
Since initial entropy of code section, data section, resource section, entire PE file, and entrypoint sections are already present in the static features, adding these novel features constitute an approximation of the entropy pattern generally studied in literature. 

To extract these features each sample was executed once for each of their executable sections. The execution continues until control flow reached the first jump instruction into the target executable section. This stop condition reached, entropy values of all distinct sections and global entropy of the file are recorded and the process is reiterated for the subsequent executable section. This way of proceeding is less powerful than monitoring all entropy changes and stopping accordingly (e.g.~a packer using multiple layers could still be engaged in the unpacking process). However, it could be applied easily to non-packed executables required in the context of packing detection.

The dynamic features were extracted for a subset of 12,036 samples from \dsthree and then all 11 ML algorithms were trained and tested using 10-fold cross-validation with $90\%$ of the data set used for training and $10\%$ for validation (as in all the previous experiments).

The impact of these features on accuracy results can be observed in Table \ref{Tab:perf_dynamic_feature}. 
These results indicate that (our) dynamic features do \emph{not} significantly influence the decision process.
At best there is a negligible increase in accuracy (e.g.~for \gnbc, \dl, and \mlp) and at worse a degradation (\svm), although most did not benefit from the addition of dynamic features.

Due to the significant extraction cost (in time, instrumentation, execution environment, etc.) of dynamic features and the lack of impact, these results show that static features can be preferred for efficient packer classification as they already provide high accuracy for packing detection.
Note that since these results indicated a negligible improvement on a restricted data set. Thus, further exploration of these features was not considered here.
This result agrees with what has been reported in literature, in that dynamic features can contribute to accurate detection, but are not necessary. Particularly when handling only detection, and when the accuracy is already very high.
Generally, when dynamic features are used in the literature, the objective is not just packing detection but also classification or direct unpacking by trying to find the original entrypoint. 
Thus the high costs of computing dynamic features need to be countered balance by stronger objectives than just malware detection. However, since IAT table features show significant influence on classification, focusing on these kinds of dynamic features could deliver more impact (similar to~\cite{DBLP:conf/ccs/Cheng0FPCZM18}). Even though overhead would generally stay important. We leave this question for future works.

\section{Conclusion}
\label{sec:Conclusion}

The challenge of detecting packed programs is a key part of contemporary cyber-security defense.
Many approaches have been taken to detect whether a sample is packed, applying various algorithms to many different features and measuring the effectiveness in different ways.
This work addresses three main research questions about the:
significant features,
effective ML algorithms, and
long term effectiveness for packing detection.

The feature significance shows that across 11 different ML algorithms there are a small number of static features that are most significant.
These are related to:
the stack reserve (Feature 19),
number of readable and writable sections (Feature 29),
non-standard entry point section (Feature 35),
raw and virtual data size of sections (Features 37 \& 39),
byte entropy of .text sections (Feature 43),
12th entry point byte (Feature 60), and
malicious API calls (Features 115-116).
This diversity of features correlates with prior results showing that many different kinds of features can be effective, and that none is clearly dominant.
Further, by gathering feature use data 
across many ML algorithms there is evidence that many features can play a small role in improving accuracy, even if they are not significant alone.

The effectiveness on 11 different ML algorithms was considered with various metrics and also considering permutations in the training data.
Overall \knn, \gbdt, \mlp, and \svm proved to be the most effective and robust.
These 4 algorithms were all able to score well on: accuracy, precision, recall, and $F$-score taking into account packed unpacked, and averaging.
There 4 algorithms also performed well in adapting to new training data effectively.

The long term effectiveness of the 11 ML algorithms was also evaluated using chronological samples.
The results here showed that all classifiers had reduced accuracy over time (as expected).
The uptime between retraining for the different algorithms varied substantially, with \knn, \rf, \gbdt, \mlp, and \svm all being able to maintain a high efficiency for $\sim 80$ days or more. However, the extremely low training cost for \dt and \rf makes them also competitive if time to retrain is weighted significantly higher than time between retraining.

Overall the results here indicate that \knn, \gbdt, \mlp, and \svm are all effective and economical. Of these 4 the choice for a particular application depends on which metrics are most important, with \knn and \gbdt scoring higher in economic ratio due to lower training time and needing less features, but tending to be lower in effectiveness.
By contrast \svm scores higher in effectiveness metrics at the cost of using more features and taking more training time.
\mlp does not stand out significantly in any one area, but has a different balance that may be optimal for some applications.

\medskip

\noindent\textbf{Acknowledgments.} Charles-Henry Bertrand Van Ouytsel is 
FRIA grantee of the Belgian Fund for
Scientific Research (FNRS-F.R.S.). We would like to thanks Cisco for their malware feed and the availability of their packing detector engine.

\medskip

\bibliographystyle{abbrv}
\bibliography{refs}

\begin{thebibliography}{10}

\bibitem{ClamAV}
{ClamAV}.
\newblock \url{https://www.clamav.net}. November 2019.

\bibitem{die}
Detect-it-easy version 2.06.
\newblock \url{https://github.com/horsicq/Detect-It-Easy}. November 2019.

\bibitem{logreg1}
Generalized linear model ({GLM}) — h2o 3.30.0.4 documentation.
\newblock
  \url{https://docs.h2o.ai/h2o/latest-stable/h2o-docs/data-science/glm.html}.

\bibitem{AVTest}
Malware statistics trends report: Av-test.
\newblock \url{https://portal.av-atlas.org/malware}. 14 January 2021.

\bibitem{peid}
{PEiD}, version 0.95.
\newblock \url{https://appnee.com/peid/}. November 2019.

\bibitem{DL8.5}
pydl8.5.
\newblock \url{https://github.com/aia-uclouvain/pydl8.5}. May 2020.

\bibitem{Shadow}
The shadowserver foundation.
\newblock \url{https://www.shadowserver.org/}. November 2019.

\bibitem{Wildlist}
The wildlist organization international.
\newblock \url{http://www.wildlist.org/}.

\bibitem{aghakhani2020malware}
H.~Aghakhani, F.~Gritti, F.~Mecca, M.~Lindorfer, S.~Ortolani, D.~Balzarotti,
  G.~Vigna, and C.~Kruegel.
\newblock When malware is packin'heat; limits of machine learning classifiers
  based on static analysis features.
\newblock In {\em Network and Distributed Systems Security (NDSS) Symposium
  2020}, 2020.

\bibitem{DBLP:conf/codaspy/AhmadiUSTG16}
M.~Ahmadi, D.~Ulyanov, S.~Semenov, M.~Trofimov, and G.~Giacinto.
\newblock Novel feature extraction, selection and fusion for effective malware
  family classification.
\newblock In {\em Proceedings of the Sixth {ACM} on Conference on Data and
  Application Security and Privacy, {CODASPY} 2016, New Orleans, LA, USA, March
  9-11, 2016}, pages 183--194, 2016.

\bibitem{peframe}
G.~Amato.
\newblock peframe version 6.0.3.
\newblock \url{https://github.com/guelfoweb/peframe}. November 2019.

\bibitem{arora2013heuristics}
R.~Arora, A.~Singh, H.~Pareek, and U.~R. Edara.
\newblock A heuristics-based static analysis approach for detecting packed pe
  binaries.
\newblock {\em International Journal of Security and Its Applications},
  7(5):257--268, 2013.

\bibitem{baldangombo_static_2013}
U.~Baldangombo, N.~Jambaljav, and S.-J. Horng.
\newblock A static malware detection system using data mining methods.

\bibitem{DBLP:journals/corr/abs-1802-10172}
R.~Balestriero, H.~Glotin, and R.~G. Baraniuk.
\newblock Semi-supervised learning enabled by multiscale deep neural network
  inversion.
\newblock {\em CoRR}, abs/1802.10172, 2018.

\bibitem{bat2013dynamic}
M.~Bat-Erdene, T.~Kim, H.~Li, and H.~Lee.
\newblock Dynamic classification of packing algorithms for inspecting
  executables using entropy analysis.
\newblock In {\em 2013 8th International Conference on Malicious and Unwanted
  Software:" The Americas"(MALWARE)}, pages 19--26. IEEE, 2013.

\bibitem{bat2017packer}
M.~Bat-Erdene, T.~Kim, H.~Park, and H.~Lee.
\newblock Packer detection for multi-layer executables using entropy analysis.
\newblock {\em Entropy}, 19(3):125, 2017.

\bibitem{bat2017entropy}
M.~Bat-Erdene, H.~Park, H.~Li, H.~Lee, and M.-S. Choi.
\newblock Entropy analysis to classify unknown packing algorithms for malware
  detection.
\newblock {\em International Journal of Information Security}, 16(3):227--248,
  2017.

\bibitem{biondi2019effective}
F.~Biondi, M.~A. Enescu, T.~Given-Wilson, A.~Legay, L.~Noureddine, and
  V.~Verma.
\newblock Effective, efficient, and robust packing detection and
  classification.
\newblock 85:436--451.

\bibitem{margaria_tutorial_2018}
F.~Biondi, T.~Given-Wilson, A.~Legay, C.~Puodzius, and J.~Quilbeuf.
\newblock Tutorial: An overview of malware detection and evasion techniques.
\newblock In T.~Margaria and B.~Steffen, editors, {\em Leveraging Applications
  of Formal Methods, Verification and Validation. Modeling}, volume 11244,
  pages 565--586. Springer International Publishing.
\newblock Series Title: Lecture Notes in Computer Science.

\bibitem{DBLP:conf/ccs/BonfanteFMRST15}
G.~Bonfante, J.~M. Fernandez, J.~Marion, B.~Rouxel, F.~Sabatier, and
  A.~Thierry.
\newblock Codisasm: Medium scale concatic disassembly of self-modifying
  binaries with overlapping instructions.
\newblock In {\em Proceedings of the 22nd {ACM} {SIGSAC} Conference on Computer
  and Communications Security, Denver, CO, USA, October 12-16, 2015}, pages
  745--756. {ACM}, 2015.

\bibitem{DBLP:conf/ccs/Cheng0FPCZM18}
B.~Cheng, J.~Ming, J.~Fu, G.~Peng, T.~Chen, X.~Zhang, and J.~Marion.
\newblock Towards paving the way for large-scale windows malware analysis:
  Generic binary unpacking with orders-of-magnitude performance boost.
\newblock In {\em Proceedings of the 2018 {ACM} {SIGSAC} Conference on Computer
  and Communications Security, {CCS} 2018, Toronto, ON, Canada, October 15-19,
  2018}, pages 395--411. {ACM}, 2018.

\bibitem{choi2008pe}
Y.-s. Choi, I.-k. Kim, J.-t. Oh, and J.-c. Ryou.
\newblock Pe file header analysis-based packed pe file detection technique
  (phad).
\newblock In {\em International Symposium on Computer Science and its
  Applications}, pages 28--31. IEEE, 2008.

\bibitem{han2009packed}
S.~Han, K.~Lee, and S.~Lee.
\newblock Packed pe file detection for malware forensics.
\newblock In {\em 2009 2nd International Conference on Computer Science and Its
  Applications, CSA 2009}, page 5404211, 2009.

\bibitem{jeong2010generic}
G.~Jeong, E.~Choo, J.~Lee, M.~Bat-Erdene, and H.~Lee.
\newblock Generic unpacking using entropy analysis.
\newblock In {\em 2010 5th International Conference on Malicious and Unwanted
  Software}, pages 98--105. IEEE, 2010.

\bibitem{kang2007renovo}
M.~G. Kang, P.~Poosankam, and H.~Yin.
\newblock Renovo: A hidden code extractor for packed executables.
\newblock In {\em Proceedings of the 2007 ACM workshop on Recurring malcode},
  pages 46--53, 2007.

\bibitem{manalyze}
I.~Kwiatkowski.
\newblock Manalyze.
\newblock \url{https://github.com/JusticeRage/Manalyze}. November 2019.

\bibitem{DBLP:journals/ieeesp/LydaH07}
R.~Lyda and J.~Hamrock.
\newblock Using entropy analysis to find encrypted and packed malware.
\newblock {\em {IEEE} Security {\&} Privacy}, 5(2):40--45, 2007.

\bibitem{ML}
A.~C. Müller and S.~Guido.
\newblock {\em Introduction to machine learning with Python: a guide for data
  scientists}.
\newblock O'Reilly.

\bibitem{nijssen2020learning}
S.~Nijssen, P.~Schaus, et~al.
\newblock Learning optimal decision trees using caching branch-and-bound
  search.
\newblock In {\em Thirty-Fourth AAAI Conference on Artificial Intelligence},
  2020.

\bibitem{noureddine2021se}
L.~Noureddine, A.~Heuser, C.~Puodzius, and O.~Zendra.
\newblock Se-pac: A self-evolving packer classifier against rapid packers
  evolution.
\newblock In {\em CODASPY'21: Eleventh ACM Conference on Data and Application
  Security and Privacy}, 2021.

\bibitem{scikit-learn}
F.~Pedregosa, G.~Varoquaux, A.~Gramfort, V.~Michel, B.~Thirion, O.~Grisel,
  M.~Blondel, P.~Prettenhofer, R.~Weiss, V.~Dubourg, J.~Vanderplas, A.~Passos,
  D.~Cournapeau, M.~Brucher, M.~Perrot, and E.~Duchesnay.
\newblock Scikit-learn: Machine learning in {P}ython.
\newblock {\em Journal of Machine Learning Research}, 12:2825--2830, 2011.

\bibitem{perdisci_classification_2008}
R.~Perdisci, A.~Lanzi, and W.~Lee.
\newblock Classification of packed executables for accurate computer virus
  detection.
\newblock 29(14):1941--1946.

\bibitem{perdisci2008mcboost}
R.~Perdisci, A.~Lanzi, and W.~Lee.
\newblock Mcboost: Boosting scalability in malware collection and analysis
  using statistical classification of executables.
\newblock In {\em 2008 Annual Computer Security Applications Conference
  (ACSAC)}, pages 301--310. IEEE, 2008.

\bibitem{raphel2015information}
J.~Raphel and P.~Vinod.
\newblock Information theoretic method for classification of packed and encoded
  files.
\newblock In {\em Proceedings of the 8th International Conference on Security
  of Information and Networks}, pages 296--303, 2015.

\bibitem{santos2011collective}
I.~Santos, X.~Ugarte-Pedrero, B.~Sanz, C.~Laorden, and P.~G. Bringas.
\newblock Collective classification for packed executable identification.
\newblock In {\em Proceedings of the 8th Annual Collaboration, Electronic
  messaging, Anti-Abuse and Spam Conference}, pages 23--30, 2011.

\bibitem{sun2010pattern}
L.~Sun, S.~Versteeg, S.~Bozta{\c{s}}, and T.~Yann.
\newblock Pattern recognition techniques for the classification of malware
  packers.
\newblock In {\em Australasian Conference on Information Security and Privacy},
  pages 370--390. Springer, 2010.

\bibitem{ugarte2011structural}
X.~Ugarte-Pedrero, I.~Santos, and P.~G. Bringas.
\newblock Structural feature based anomaly detection for packed executable
  identification.
\newblock In {\em Computational intelligence in security for information
  systems}, pages 230--237. Springer, 2011.

\bibitem{ugarte2014adoption}
X.~Ugarte-Pedrero, I.~Santos, I.~Garc{\'\i}a-Ferreira, S.~Huerta, B.~Sanz, and
  P.~G. Bringas.
\newblock On the adoption of anomaly detection for packed executable filtering.
\newblock {\em Computers \& Security}, 43:126--144, 2014.

\bibitem{ugarte2012countering}
X.~Ugarte-Pedrero, I.~Santos, B.~Sanz, C.~Laorden, and P.~G. Bringas.
\newblock Countering entropy measure attacks on packed software detection.
\newblock In {\em 2012 IEEE Consumer Communications and Networking Conference
  (CCNC)}, pages 164--168. IEEE, 2012.

\bibitem{YARA}
VirusTotal.
\newblock Virustotal: Yara in a nutshell (2019).

\bibitem{zakeri2015static}
M.~Zakeri, F.~Faraji~Daneshgar, and M.~Abbaspour.
\newblock A static heuristic approach to detecting malware targets.
\newblock {\em Security and Communication Networks}, 8(17):3015--3027, 2015.

\end{thebibliography}

\newpage
\newpage

\begin{appendix}

\section{Extracted Features}
\label{sec:appendix_extracted_features}

This section includes a full table (\ref{tab:full_features_table}) of all the 119 static features considered in this work and their descriptions. They are also grouped according to their commonality as in Biondi et al.~\cite{biondi2019effective}.

\begin{longtable}{|l|l|p{12cm}|}
    \hline
    \textbf{ID} & \textbf{Type} & \textbf{Description} \\
        \hline
    \multicolumn{3}{|c|}{\textbf{Metadata features}} \\
    \hline \hline
    1 & \textit{Boolean} & DLL can be relocated at load time. \textit{Extracted from DLLs characteristics} \\
    \hline
    2 & \textit{Boolean} & Code Integrity checks are enforced. \textit{Extracted from DLLs characteristics} \\
    \hline
    3 & \textit{Boolean} & Image is NX compatible. \textit{Extracted from DLLs characteristics} \\
    \hline
    4 & \textit{Boolean} & Isolation aware, but do not isolate the image. \textit{Extracted from DLLs characteristics} \\
    \hline
    5 & \textit{Boolean} & Does not use structured exception (SE) handling. No SE handler may be called in this image.  \textit{Extracted from DLLs characteristics} \\
    \hline
    6 & \textit{Boolean} & Do not bind the image. \textit{Extracted from DLLs characteristics} \\
    \hline
    7 & \textit{Boolean} & A WDM driver.  \textit{Extracted from DLLs characteristics} \\
    \hline
    8 & \textit{Boolean} & Terminal Server aware. \textit{Extracted from DLLs characteristics} \\
    \hline
    9 & \textit{Integer} & The image file checksum.  \\
    \hline
    10 & \textit{Integer} & The preferred address of the first byte of image when loaded into memory \\
    \hline
    11 & \textit{Integer} & The address that is relative to the image base of the beginning-of-code section when it is loaded into memory.\\
    \hline
    12 & \textit{Integer} & The major version number of the required operating system.\\
    \hline
    13 & \textit{Integer} & The minor version number of the required operating system.\\
    \hline
    14 & \textit{Integer} & The size (in bytes) of the image, including all headers, as the image is loaded in memory.\\
    \hline
    15 & \textit{Integer} & The size of the code (.text) section, or the sum of all code sections if there are multiple sections.\\
    \hline
    16 & \textit{Integer} & The combined size of an MS DOS stub, PE header, and section headers rounded up to a multiple of FileAlignment.\\
    \hline
    17 & \textit{Integer} & The size of the initialized data section, or the sum of all such sections if there are multiple data sections.\\
    \hline
    18 & \textit{Integer} & The size of the uninitialized data section (BSS), or the sum of all such sections if there are multiple BSS sections \\
    \hline
    19 & \textit{Integer} & The size of the stack to reserve.\\
    \hline
    20 & \textit{Integer} & The size of the stack to commit.\\
    \hline
    21 & \textit{Integer} & The alignment (in bytes) of sections when they are loaded into memory.\\
    \hline \hline
    \multicolumn{3}{|c|}{\textbf{Section features}}\\
    \hline \hline
    22 & \textit{Integer} & The number of standard sections the PE holds\\
    \hline
    23 & \textit{Integer} & The number of non-standard sections the PE holds\\
    \hline
    24 & \textit{Float} & The ratio between the number of standard sections found and the number of all sections found in the PE under analysis\\
    \hline
    25 & \textit{Integer} & The number of Executable sections the PE holds\\
    \hline
    26 & \textit{Integer} & The number of Writable sections the PE holds\\
    \hline
    27 & \textit{Integer} & The number of Writable and Executable sections the PE holds\\
    \hline
    28 & \textit{Integer} & The number of readable and executable sections\\
    \hline
    29 & \textit{Integer} & The number of readable and writable sections\\
    \hline
    30 & \textit{Integer} & The number of Writable and Readable and Executable sections the PE holds\\
    \hline
    31 & \textit{Boolean} & The code section is not executable\\
    \hline
    32 & \textit{Boolean} & The executable section is not a code section\\
    \hline
    33 & \textit{Boolean} & The code section is not present in the PE under analysis\\
    \hline
    34 & \textit{Boolean} & The entry point is not in the code section\\
    \hline
    35 & \textit{Boolean} & The entry point is not in a standard section\\
    \hline
    36 & \textit{Boolean} & The entry point is not in an executable section\\
    \hline
    37 & \textit{Float} & The ratio between raw data and virtual size for the section of the entry point\\
    \hline
    38 & \textit{Integer} & The number of section having their raw data size zero\\
    \hline
    39 & \textit{Integer} & The number of sections having their virtual size greater than their raw data size.\\
    \hline
    40 & \textit{Float} & The maximum ratio raw data to virtual size among all sections\\
    \hline
    41 & \textit{Float} & the minimum ratio raw data to virtual size among all sections\\
    \hline
    42 & \textit{Boolean} & The address pointing to raw data on disk is not conforming with the file alignment\\
    \hline
     \multicolumn{3}{|c|}{\textbf{Entropy features}}\\    
    \hline \hline
    43 & \textit{Float $\in [0;8]$} & The byte entropy of code (.text) sections\\
    \hline
    44 & \textit{Float $\in [0;8]$} & The byte entropy of data section\\
    \hline
    45 & \textit{Float $\in [0;8]$} & The byte entropy of resource section\\
    \hline
    46 & \textit{Float $\in [0;8]$} & The byte entropy of PE header\\
    \hline
    47 & \textit{Float $\in [0;8]$} & The byte entropy of the entire PE file\\
    \hline
    48 & \textit{Float $\in [0;8]$} & The byte entropy of the section holding the entry point of the PE under analysis\\
    \hline
    \multicolumn{3}{|c|}{\textbf{Entry byte features}} \\    
    \hline \hline
    49 - 112 & \textit{Integer $\in [0;255]$} & Values of 64 bytes following the entry point, each byte correspond to 1 feature position\\
    \hline
    \multicolumn{3}{|c|}{\textbf{Import functions features}}\\    
    \hline \hline
    113 & \textit{Integer} & The number of DLLs imported\\
    \hline
    114 & \textit{Integer} & The number of functions imported found in the import table directory (IDT)\\
    \hline
    115 & \textit{Integer} & The number of malicious APIs imported (malicious as defined in~\cite{zakeri2015static}). This list includes: "GetProcAddress", "LoadLibraryA", "LoadLibrary", "ExitProcess", "GetModuleHandleA", "VirtualAlloc", "VirtualFree", "GetModuleFileNameA", "CreateFileA", "RegQueryValueExA", "MessageBoxA", "GetCommandLineA", "VirtualProtect", "GetStartupInfoA", "GetStdHandle", "RegOpenKeyExA". \\
    \hline
    116 & \textit{Float} & The ratio between the number of malicious APIs imported to the number of all functions imported by the PE\\
    \hline
    117 & \textit{Integer} & the number of addresses (corresponds to functions) found in the import address table (IAT)\\
\hline
    \multicolumn{3}{|c|}{\textbf{Ressource features}} \\    
    
    \hline \hline
    118 & \textit{Boolean} & The debug directory is present or not\\
    \hline
    119 & \textit{Integer} & The number of resources the PE holds\\
    
    \hline
    \caption{Features description.}
    \label{tab:full_features_table}
    \end{longtable}

\section{Boolean Conversion Applied to Features}
\label{sec:appendix_boolean_conversion}

This section details in Table~\ref{tab:full_boolean_conversion} how each of the 119 features (detailed in Table~\ref{tab:full_features_table}) is converted into Booleans via bucketing.

\begin{longtable}{|l|p{12cm}|}
    \hline
    \textbf{ID}  & \textbf{Description} \\
    \hline
    1-8  & Already boolean \\
    \hline
    9  & 2 buckets : 0 or other values \\
    \hline
    10 & 2 buckets : 4194304 or other values \\
    \hline
    11  & 3 buckets : 4096, 8192 or other values\\
    \hline
    12 & 3 buckets : 4, 5 or other values \\
    \hline
    13 & 2 buckets : 0 or other values\\
    \hline
    14 &  2 buckets : $<250000$ or $>=25000$\\
    \hline
    15 & 2 buckets : $<50000$ or $>=50000$ \\
    \hline
    16 & 5 buckets : 1024, 4096, 512, 1536 or other values \\
    \hline
    17 & feature deleted due to high sparsity \\
    \hline
    18 & 2 buckets : 0 or other values\\
    \hline
    19  & 2 buckets : 1048576 or other values\\
    \hline
    20 & 5 buckets : 4096, 16384, 8192, 65536 or other values \\
    \hline
    21 & 3 buckets : 4096, 8192 or other values\\
    \hline
    22 &  10 buckets : 0, 1, 2, 3, 4, 5, 6, 7, 8 or $>=9$\\
    \hline
    23 & 5 buckets : 0, 1, 2, 3 or $>=4$ \\
    \hline
    24 & 2 buckets : 0 or other values\\
    \hline
    25 & 2 buckets : $<3$ or $>=3$ \\
    \hline
    26 & 2 buckets : 0 or other values\\
    \hline
    23 & 5 buckets : 0, 1, 2, 3 or $>=4$ \\
    \hline
    24 & 2 buckets : 0 or other values\\
    \hline
    25 & 2 buckets : $<3$ or $>=3$ \\
    \hline
    26 & 6 buckets : 0, 1, 2, 3, 4 or $>=5$\\
    \hline
        27 & 4 buckets : 0, 1, 2 or $>=3$\\
    \hline
        28 & 4 buckets : 0, 1, 2 or $>=3$\\
    \hline
            29 & 6 buckets : 0, 1, 2, 3, 4 or $>=5$\\
    \hline
            30 & 4 buckets : 0, 1, 2 or $>=3$\\
    \hline
    31-36  & Already boolean \\    
    \hline
    37 & 3 buckets : $>1$, $<1$ or 1\\
    \hline
    38 & 4 buckets : 0, 1, 2 or $>=3$\\
    \hline
    39 & 5 buckets : 0, 1, 2, 3 or $>=4$\\
    \hline
    40 & 3 buckets : $>1$, $<1$ or 1\\
    \hline
    41 & 2 buckets : 0 or other values\\
    \hline
    42  & Already boolean \\   
    \hline
    43-45  & 2 buckets : $-1$ or other values \\  
    \hline
    46-117  & 2 buckets : 0 or other values \\  
    \hline
    118  & Already boolean \\  
    \hline
    119  & 2 buckets : 0 or other values \\  
    \hline
    \caption{Explanation of Boolean conversion with bucketing for all 119 features.}
    \label{tab:full_boolean_conversion}
    \end{longtable}
 \newpage

\end{appendix}

\end{document}